\documentclass[11pt, a4paper]{article}
\PassOptionsToPackage{unicode}{hyperref}
\usepackage{amssymb}
\usepackage{jheppub}
\usepackage{upgreek}
\usepackage{slashed}
\usepackage{tikz}
\usepackage[caption=false,labelformat=simple]{subfig}

\title{Wilson--'t Hooft lines as transfer matrices}

\author[a]{Kazunobu Maruyoshi,}
\author[b, c]{Toshihiro Ota}
\author[d, e]{and Junya Yagi}

\affiliation[a]{Faculty of Science and Technology, Seikei University\\
  3--3--1 Kichijoji-Kitamachi, Musashino-shi, Tokyo 180--8633 Japan}

\affiliation[b]{Department of Physics, Osaka University \\
  Toyonaka, Osaka 560--0043 Japan}

\affiliation[c]{interdisciplinary Theoretical \& Mathematical Sciences Program (iTHEMS), RIKEN \\
  Wako, Saitama 351--0198 Japan}

\affiliation[d]{Perimeter Institute for Theoretical Physics \\
  Waterloo, ON, N2L 2Y5 Canada}

\affiliation[e]{Yau Mathematical Sciences Center, Tsinghua University \\
  Beijing 100084, P.R. China}

\emailAdd{maruyoshi@st.seikei.ac.jp}
\emailAdd{tota@het.phys.sci.osaka-u.ac.jp}
\emailAdd{junyagi@tsinghua.edu.cn}


\arxivnumber{}

\let\U\relax
\let\C\relax

\newcommand{\gf}{\mathfrak{g}}

\newcommand{\hf}{\mathfrak{h}}

\newcommand{\tf}{\mathfrak{t}}

\newcommand{\id}{\mathop{\mathrm{id}}\nolimits}

\def\ie{\begin{equation}\begin{aligned}}
\def\fe{\end{aligned}\end{equation}}

\newcommand{\Bigket}[1]{\Bigl|#1\Bigr\rangle}
\newcommand{\Bigbra}[1]{\Bigl\langle #1\Bigr|}

\newcommand{\vev}[1]{\langle #1 \rangle}

\newcommand{\biggvev}[1]{\biggl\langle #1 \biggr\rangle}

\newcommand{\Hom}{\mathop{\mathrm{Hom}}\nolimits}

\renewcommand{\Im}{\mathop{\mathrm{Im}}\nolimits}
\renewcommand{\Re}{\mathop{\mathrm{Re}}\nolimits}

\newcommand{\Tr}{\mathop{\mathrm{Tr}}\nolimits}
\newcommand{\End}{\mathop{\mathrm{End}}\nolimits}

\newcommand{\SU}{\mathrm{SU}}
\newcommand{\PSU}{\mathrm{PSU}}

\newcommand{\glf}{\mathfrak{gl}}
\newcommand{\slf}{\mathfrak{sl}}
\newcommand{\suf}{\mathfrak{su}}

\newcommand{\U}{\mathrm{U}}

\newcommand{\iso}{\cong}

\newcommand{\Z}{\mathbb{Z}}

\newcommand{\R}{\mathbb{R}}
\newcommand{\C}{\mathbb{C}}

\let\nc\newcommand
\let\renc\renewcommand
\nc{\wbar}{\overline}
\let\td\tilde
\let\wtd\widetilde
\let\wht\widehat
\let\mcl\mathcal

\nc{\ab}{{\bar{a}}} \nc{\at}{\tilde{a}} \nc{\ah}{\hat{a}}
\nc{\bb}{{\bar{b}}} \nc{\bt}{\tilde{b}} \nc{\bh}{\hat{b}}
\nc{\cb}{{\bar{c}}} \nc{\ct}{\tilde{c}} 
\nc{\db}{{\bar{d}}} \nc{\dt}{\tilde{d}} \renc{\dh}{\hat{d}}
\nc{\eb}{{\bar{e}}} \nc{\et}{\tilde{e}} \nc{\eh}{\hat{e}}
\nc{\fb}{{\bar{f}}} \nc{\ft}{\tilde{f}} \nc{\fh}{\hat{f}}
\nc{\gb}{{\bar{g}}} \nc{\gt}{\tilde{g}} \nc{\gh}{\hat{g}}
\nc{\hb}{{\bar{h}}} \nc{\hh}{\hat{h}} 
\nc{\ib}{{\bar{\imath}}} \nc{\ih}{\hat{\imath}} 
\nc{\jb}{{\bar{\jmath}}} \nc{\jt}{\tilde{\jmath}} \nc{\jh}{\hat{\jmath}}
\nc{\kb}{{\bar{k}}} \nc{\kt}{\tilde{k}} \nc{\kh}{\hat{k}}
\nc{\lb}{{\bar{l}}} \nc{\lt}{\tilde{l}} \nc{\lh}{\hat{l}}
\nc{\mb}{{\bar{m}}} \nc{\mt}{\tilde{m}} \nc{\mh}{\hat{m}}
\nc{\nb}{{\bar{n}}} \nc{\nt}{\tilde{n}} \nc{\nh}{\hat{n}}
\nc{\ob}{{\bar{o}}} \nc{\ot}{\tilde{o}} \nc{\oh}{\hat{o}}
\nc{\pb}{{\bar{p}}} \nc{\pt}{\tilde{p}} \nc{\ph}{\hat{p}}
\nc{\qb}{{\bar{q}}} \nc{\qt}{\tilde{q}} \nc{\qh}{\hat{q}}
\nc{\rb}{{\bar{r}}} \nc{\rt}{\tilde{r}} \nc{\rh}{{\hat{r}}}
\renc{\sb}{{\bar{s}}} \nc{\st}{\tilde{s}} \nc{\sh}{\hat{s}}
\nc{\tb}{{\bar{t}}} \renc{\th}{\hat{t}} 
\nc{\ub}{{\bar{u}}} \nc{\ut}{\tilde{u}} \nc{\uh}{\hat{u}}
\nc{\vb}{{\bar{v}}} \nc{\vt}{\tilde{v}} \nc{\vh}{\hat{v}}
\nc{\wb}{{\bar{w}}} \nc{\wt}{\tilde{w}} \nc{\wh}{\hat{w}}
\nc{\xb}{{\bar{x}}} \nc{\xt}{\tilde{x}} \nc{\xh}{\hat{x}}
\nc{\yb}{{\bar{y}}} \nc{\yt}{\tilde{y}} \nc{\yh}{\hat{y}}
\nc{\zb}{{\bar{z}}} \nc{\zt}{\tilde{z}} \nc{\zh}{\hat{z}}

\nc{\Ab}{{\wbar{A}}} \nc{\At}{{\wtd{A}}} \nc{\Ah}{{\wht{A}}}
\nc{\Bb}{{\wbar{B}}} \nc{\Bt}{{\wtd{B}}} \nc{\Bh}{{\wht{B}}}
\nc{\Cb}{{\wbar{C}}} \nc{\Ct}{{\wtd{C}}} \nc{\Ch}{{\wht{C}}}
\nc{\Db}{{\wbar{D}}} \nc{\Dt}{{\wtd{D}}} \nc{\Dh}{{\wht{D}}}
\nc{\Eb}{{\wbar{E}}} \nc{\Et}{{\wtd{E}}} \nc{\Eh}{{\wht{E}}}
\nc{\Fb}{{\wbar{F}}} \nc{\Ft}{{\wtd{F}}} \nc{\Fh}{{\wht{F}}}
\nc{\Gb}{{\wbar{G}}} \nc{\Gt}{{\wtd{G}}} \nc{\Gh}{{\wht{G}}}
\nc{\Hb}{{\wbar{H}}} \nc{\Ht}{{\wtd{H}}} \nc{\Hh}{{\wht{H}}}
\nc{\Ib}{{\bar{I}}} \nc{\It}{{\wtd{I}}} \nc{\Ih}{{\wht{I}}}
\nc{\Jb}{{\wbar{J}}} \nc{\Jt}{{\wtd{J}}} \nc{\Jh}{{\wht{J}}}
\nc{\Kb}{{\wbar{K}}} \nc{\Kt}{{\wtd{K}}} \nc{\Kh}{{\wht{K}}}
\nc{\Lb}{{\wbar{L}}} \nc{\Lt}{{\wtd{L}}} \nc{\Lh}{{\wht{L}}}
\nc{\Mb}{{\wbar{M}}} \nc{\Mt}{{\wtd{M}}} \nc{\Mh}{{\wht{M}}}
\nc{\Nb}{{\wbar{N}}} \nc{\Nt}{{\wtd{N}}} \nc{\Nh}{{\wht{N}}}
\nc{\Ob}{{\wbar{O}}} \nc{\Ot}{{\wtd{O}}} \nc{\Oh}{{\wht{O}}}
\nc{\Pb}{{\wbar{P}}} \nc{\Pt}{{\wtd{P}}} \nc{\Ph}{{\wht{P}}}
\nc{\Qb}{{\wbar{Q}}} \nc{\Qt}{{\wtd{Q}}} \nc{\Qh}{{\wht{Q}}}
\nc{\Rb}{{\wbar{R}}} \nc{\Rt}{{\wtd{R}}} \nc{\Rh}{{\wht{R}}}
\nc{\Sb}{{\wbar{S}}} \nc{\St}{{\wtd{S}}} \nc{\Sh}{{\wht{S}}}
\nc{\Tb}{{\wbar{T}}} \nc{\Tt}{{\wtd{T}}} \nc{\Th}{{\wht{T}}}
\nc{\Ub}{{\wbar{U}}} \nc{\Ut}{{\wtd{U}}} \nc{\Uh}{{\wht{U}}}
\nc{\Vb}{{\wbar{V}}} \nc{\Vt}{{\wtd{V}}} \nc{\Vh}{{\wht{V}}}
\nc{\Wb}{{\wbar{W}}} \nc{\Wt}{{\wtd{W}}} \nc{\Wh}{{\wht{W}}}
\nc{\Xb}{{\wbar{X}}} \nc{\Xt}{{\wtd{X}}} \nc{\Xh}{{\wht{X}}}
\nc{\Yb}{{\wbar{Y}}} \nc{\Yt}{{\wtd{Y}}} \nc{\Yh}{{\wht{Y}}}
\nc{\Zb}{{\wbar{Z}}} \nc{\Zt}{{\wtd{Z}}} \nc{\Zh}{{\wht{Z}}}

\nc{\CA}{{\mcl{A}}} \nc{\CAb}{{\wbar{\CA}}} \nc{\CAt}{{\wtd{\CA}}} \nc{\CAh}{{\wht{\CA}}}
\nc{\CB}{{\mcl{B}}} \nc{\CBb}{{\wbar{\CB}}} \nc{\CBt}{{\wtd{\CB}}} \nc{\CBh}{{\wht{\CB}}}
\nc{\CC}{{\mcl{C}}} \nc{\CCb}{{\wbar{\CC}}} \nc{\CCt}{{\wtd{\CC}}} \nc{\CCh}{{\wht{\CC}}}
\nc{\cD}{{\mcl{D}}} \nc{\cDb}{{\wbar{\cD}}} \nc{\cDt}{{\wtd{\cC}}} \nc{\cDh}{{\wht{\cD}}}
\nc{\CE}{{\mcl{E}}} \nc{\CEb}{{\wbar{\CE}}} \nc{\CEt}{{\wtd{\CE}}} \nc{\CEh}{{\wht{\CE}}}
\nc{\CF}{{\mcl{F}}} \nc{\CFb}{{\wbar{\CF}}} \nc{\CFt}{{\wtd{\CF}}} \nc{\CFh}{{\wht{\CF}}}
\nc{\CG}{{\mcl{G}}} \nc{\CGb}{{\wbar{\CG}}} \nc{\CGt}{{\wtd{\CG}}} \nc{\CGh}{{\wht{\CG}}}
\nc{\CH}{{\mcl{H}}} \nc{\CHb}{{\wbar{\CH}}} \nc{\CHt}{{\wtd{\CH}}} \nc{\CHh}{{\wht{\CH}}}
\nc{\CI}{{\mcl{I}}} \nc{\CIb}{{\wbar{\CI}}} \nc{\CIt}{{\wtd{\CI}}} \nc{\CIh}{{\wht{\CI}}}
\nc{\CJ}{{\mcl{J}}} \nc{\CJb}{{\wbar{\CJ}}} \nc{\CJt}{{\wtd{\CJ}}} \nc{\CJh}{{\wht{\CJ}}}
\nc{\CK}{{\mcl{K}}} \nc{\CKb}{{\wbar{\CK}}} \nc{\CKt}{{\wtd{\CK}}} \nc{\CKh}{{\wht{\CK}}}
\nc{\CL}{{\mcl{L}}} \nc{\CLb}{{\wbar{\CL}}} \nc{\CLt}{{\wtd{\CL}}} \nc{\CLh}{{\wht{\CL}}}
\nc{\CM}{{\mcl{M}}} \nc{\CMb}{{\wbar{\CM}}} \nc{\CMt}{{\wtd{\CM}}} \nc{\CMh}{{\wht{\CM}}}
\nc{\CN}{{\mcl{N}}} \nc{\CNb}{{\wbar{\CN}}} \nc{\CNt}{{\wtd{\CN}}} \nc{\CNh}{{\wht{\CN}}}
\nc{\CO}{{\mcl{O}}} \nc{\COb}{{\wbar{\CO}}} \nc{\COt}{{\wtd{\CO}}} \nc{\COh}{{\wht{\CO}}}
\nc{\CP}{{\mcl{P}}} \nc{\CPb}{{\wbar{\CP}}} \nc{\CPt}{{\wtd{\CP}}} \nc{\CPh}{{\wht{\CP}}}
\nc{\CQ}{{\mcl{Q}}} \nc{\CQb}{{\wbar{\CQ}}} \nc{\CQt}{{\wtd{\CQ}}} \nc{\CQh}{{\wht{\CQ}}}
\nc{\CR}{{\mcl{R}}} \nc{\CRb}{{\wbar{\CR}}} \nc{\CRt}{{\wtd{\CR}}} \nc{\CRh}{{\wht{\CR}}}
\nc{\CS}{{\mcl{S}}} \nc{\CSb}{{\wbar{\CS}}} \nc{\CSt}{{\wtd{\CS}}} \nc{\CSh}{{\wht{\CS}}}
\nc{\CT}{{\mcl{T}}} \nc{\CTb}{{\wbar{\CT}}} \nc{\CTt}{{\wtd{\CT}}} \nc{\CTh}{{\wht{\CT}}}
\nc{\CU}{{\mcl{U}}} \nc{\CUb}{{\wbar{\CU}}} \nc{\CUt}{{\wtd{\CU}}} \nc{\CUh}{{\wht{\CU}}}
\nc{\CV}{{\mcl{V}}} \nc{\CVb}{{\wbar{\CV}}} \nc{\CVt}{{\wtd{\CV}}} \nc{\CVh}{{\wht{\CV}}}
\nc{\CW}{{\mcl{W}}} \nc{\CWb}{{\wbar{\CW}}} \nc{\CWt}{{\wtd{\CW}}} \nc{\CWh}{{\wht{\CW}}}
\nc{\CX}{{\mcl{X}}} \nc{\CXb}{{\wbar{\CX}}} \nc{\CXt}{{\wtd{\CX}}} \nc{\CXh}{{\wht{\CX}}}
\nc{\CY}{{\mcl{Y}}} \nc{\CYb}{{\wbar{\CY}}} \nc{\CYt}{{\wtd{\CY}}} \nc{\CYh}{{\wht{\CY}}}
\nc{\CZ}{{\mcl{Z}}} \nc{\CZb}{{\wbar{\CZ}}} \nc{\CZt}{{\wtd{\CZ}}} \nc{\CZh}{{\wht{\CZ}}}

\let\eps\epsilon
\let\ups\upsilon
\let\veps\varepsilon
\let\vtht\vartheta

\let\vsgm\varsigma
\let\vphi\varphi
\let\vrho\varrho

\nc{\alphab}{{\bar{\alpha}}} \nc{\alphat}{{\td{\alpha}}} \nc{\alphah}{{\hat{\alpha}}}
\nc{\betab}{{\bar{\beta}}}   \nc{\betat}{{\td{\beta}}}   \nc{\betah}{{\hat{\beta}}} 
\nc{\gammab}{{\bar{\gamma}}} \nc{\gammat}{{\td{\gamma}}} \nc{\gammah}{{\hat{\gamma}}} 
\nc{\deltab}{{\bar{\delta}}} \nc{\deltat}{{\td{\delta}}} \nc{\deltah}{{\hat{\delta}}} 
\nc{\epsilonb}{{\bar{\eps}}} \nc{\epsilont}{{\td{\eps}}} \nc{\epsilonh}{{\hat{\eps}}} 
\nc{\vepsb}{{\bar{\veps}}}   \nc{\vepst}{{\td{\veps}}}   \nc{\vepsh}{{\hat{\veps}}} 
\nc{\zetab}{{\bar{\zeta}}}   \nc{\zetat}{{\td{\zeta}}}   \nc{\zetah}{{\hat{\zeta}}} 
\nc{\etab}{{\bar{\eta}}}     \nc{\etat}{{\td{\eta}}}     \nc{\etah}{{\hat{\eta}}} 
\nc{\thetab}{{\bar{\theta}}} \nc{\thetat}{{\td{\theta}}} \nc{\thetah}{{\hat{\theta}}} 
\nc{\vthetab}{{\bar{\vtht}}} \nc{\vthetat}{{\td{\vtht}}} \nc{\vthetah}{{\hat{\vtht}}} 
\nc{\lambdab}{{\bar{\lambda}}} \nc{\lambdat}{{\td{\lambda}}} \nc{\lambdah}{{\hat{\lambda}}} 
\nc{\iotab}{{\bar{\iota}}}   \nc{\iotat}{{\td{\iota}}}   \nc{\iotah}{{\hat{\iota}}} 
\nc{\kappab}{{\bar{\kappa}}} \nc{\kappat}{{\td{\kappa}}} \nc{\kappah}{{\hat{\kappa}}} 
\nc{\lmdb}{{\bar{\lmd}}}     \nc{\lmdt}{{\td{\lmd}}}     \nc{\lmdh}{{\hat{\lmd}}} 
\nc{\mub}{{\bar{\mu}}}       \nc{\mut}{{\td{\mu}}}       \nc{\muh}{{\hat{\mu}}} 
\nc{\nub}{{\bar{\nu}}}       \nc{\nut}{{\td{\nu}}}       \nc{\nuh}{{\hat{\nu}}} 
\nc{\xib}{{\bar{\xi}}}       \nc{\xit}{{\td{\xi}}}       \nc{\xih}{{\hat{\xi}}} 
\nc{\pib}{{\bar{\pi}}}       \nc{\pit}{{\td{\pi}}}       \nc{\pih}{{\hat{\pi}}} 
\nc{\vpib}{{\bar{\vpi}}}     \nc{\vpit}{{\td{\vpi}}}     \nc{\vpih}{{\hat{\vpi}}} 
\nc{\rhob}{{\bar{\rho}}}     \nc{\rhot}{{\td{\rho}}}     \nc{\rhoh}{{\hat{\rho}}} 
\nc{\vrhob}{{\bar{\vrho}}}   \nc{\vrhot}{{\td{\vrho}}}   \nc{\vrhoh}{{\hat{\vrho}}} 
\nc{\sigmab}{{\bar{\sigma}}} \nc{\sigmat}{{\td{\sigma}}} \nc{\sigmah}{{\hat{\sigma}}} 
\nc{\vsigmab}{{\bar{\vsgm}}} \nc{\vsigmat}{{\td{\vsgm}}} \nc{\vsigmah}{{\hat{\vsgm}}} 
\nc{\taub}{{\bar{\tau}}}     \nc{\taut}{{\td{\tau}}}     \nc{\tauh}{{\hat{\tau}}} 
\nc{\upsb}{{\bar{\ups}}} \nc{\upst}{{\td{\ups}}} \nc{\upsh}{{\hat{\ups}}} 
\nc{\phib}{{\bar{\phi}}}     \nc{\phit}{{\td{\phi}}}     \nc{\phih}{{\hat{\phi}}} 
\nc{\varphib}{{\bar{\vphi}}}   \nc{\varphit}{{\td{\vphi}}}   \nc{\varphih}{{\hat{\vphi}}} 
\nc{\chib}{{\bar{\chi}}}     \nc{\chit}{{\td{\chi}}}     \nc{\chih}{{\hat{\chi}}} 
\nc{\psib}{{\bar{\psi}}}     \nc{\psit}{{\td{\psi}}}     \nc{\psih}{{\hat{\psi}}} 
\nc{\omegab}{{\bar{\omega}}} \nc{\omegat}{{\td{\omega}}} \nc{\omegah}{{\hat{\omega}}} 

\nc{\Gammab}{{\wbar{\Gamma}}}     \nc{\Gammat}{{\wtd{\Gamma}}}     \nc{\Gammah}{{\wht{\Gamma}}}
\nc{\Deltab}{{\wbar{\Delta}}}     \nc{\Deltat}{{\wtd{\Delta}}}     \nc{\Deltah}{{\wht{\Delta}}}
\nc{\Thetab}{{\wbar{\Theta}}}     \nc{\Thetat}{{\wtd{\Theta}}}     \nc{\Thetah}{{\wht{\Theta}}}
\nc{\Lambdab}{{\wbar{\Lambda}}}   \nc{\Lambdat}{{\wtd{\Lambda}}}   \nc{\Lambdah}{{\wht{\Lambda}}}
\nc{\Xib}{{\wbar{\Xi}}}           \nc{\Xit}{{\wtd{\Xi}}}           \nc{\Xih}{{\wht{\Xi}}}
\nc{\Pib}{{\wbar{\Pi}}}           \nc{\Pit}{{\wtd{\Pi}}}           \nc{\Pih}{{\wht{\Pi}}}
\nc{\Sigmab}{{\wbar{\Sigma}}}     \nc{\Sigmat}{{\wtd{\Sigma}}}     \nc{\Sigmah}{{\wht{\Sigma}}}
\nc{\Upsilonb}{{\wbar{\Upsilon}}} \nc{\Upsilont}{{\wtd{\Upsilon}}} \nc{\Upsilonh}{{\wht{\Upsilon}}}
\nc{\Phib}{{\wbar{\Phi}}} \nc{\Phit}{{\wtd{\Phi}}} \nc{\Phih}{{\wht{\Phi}}}
\nc{\Psib}{{\wbar{\Psi}}}         \nc{\Psit}{{\wtd{\Psi}}}         \nc{\Psih}{{\wht{\Psi}}}
\nc{\Omegab}{{\wbar{\Omega}}}     \nc{\Omegat}{{\wtd{\Omega}}}     \nc{\Omegah}{{\wht{\Omega}}}

\newcommand{\rmd}{\mathrm{d}}

\newcommand{\iu}{\mathrm{i}}

\let\starx\star
\let\star\relax
\newcommand{\star}{\mathop{\starx}\nolimits}

\usepackage[mathscr]{euscript}

\newcommand{\as}{\mathsf{a}}
\newcommand{\bs}{\mathsf{b}}
\newcommand{\ms}{\mathsf{m}}

\newcommand{\rootl}{\Lambda_{\mathrm{r}}}
\newcommand{\corootl}{\Lambda_{\mathrm{cr}}}
\newcommand{\weightl}{\Lambda_{\mathrm{w}}}
\newcommand{\coweightl}{\Lambda_{\mathrm{cw}}}

\usetikzlibrary{calc, positioning}
\usetikzlibrary{decorations.markings, arrows, bending}
\tikzset{
  font=\footnotesize,
  >=stealth,
  ->-/.style={decoration={
      markings, mark=at position #1 with
      {\arrow{>}}},postaction={decorate}},
  -<-/.style={decoration={
      markings, mark=at position #1 with
      {\arrow{<}}},postaction={decorate}},
  spin/.style={draw, fill=white, shape=circle, minimum size=11pt, inner
    sep=0pt, font=\scriptsize},
  node/.style={draw, fill=white, shape=circle, minimum size=11pt, inner
    sep=0pt},
  gnode/.style={node},
  fnode/.style={node, shape=rectangle},
  tnode/.style={fnode, double, minimum size=12pt},
  q-/.style={-},
  q->/.style={->, shorten >=1pt, font=\smaller[2]},
  q<-/.style={q->, <-, shorten >=0pt, shorten <=1pt},
  eq-/.style={double, double distance=2pt},
}

\newcommand{\mcharge}{\mathbf{m}}
\newcommand{\echarge}{\mathbf{e}}

\abstract{We establish a correspondence between a class of Wilson--'t
  Hooft lines in four-dimensional $\CN = 2$ supersymmetric gauge
  theories described by circular quivers and transfer matrices
  constructed from dynamical L-operators for trigonometric quantum
  integrable systems.  We compute the vacuum expectation values of the
  Wilson--'t Hooft lines in a twisted product space
  $S^1 \times_\eps \R^2 \times \R$ by supersymmetric localization and
  show that they are equal to the Wigner transforms of the transfer
  matrices.  A variant of the AGT correspondence implies an
  identification of the transfer matrices with Verlinde operators in
  Toda theory, which we also verify.  We explain how these field
  theory setups are related to four-dimensional Chern--Simons theory
  via embedding into string theory and dualities.}

\begin{document}
\makeatletter
\gdef\@fpheader{}
\makeatother

\noindent
{\small
\hfill OU-HET-1067 / RIKEN-iTHEMS-Report-20}

\maketitle

\section{Introduction}
\label{sec:intro}

Supersymmetric gauge theories in four dimensions have various
interrelated connections to quantum integrable systems.  One such
connection involves a family of surface defects in a class of
$\CN = 1$ supersymmetric gauge theories, described by planer quivers.
These surface defects act on the supersymmetric indices of the
theories as commuting difference operators shifting flavor fugacities.
It turns out that the difference operators coincide with transfer
matrices of elliptic quantum integrable
systems~\cite{Bullimore:2014nla, Maruyoshi:2016caf, Yagi:2017hmj}.

In this paper we present a similar correspondence.  This
correspondence, however, is between line defects in $\CN = 2$
supersymmetric gauge theories, described by circular quivers, and
transfer matrices of trigonometric quantum integrable systems.

$\CN = 2$ supersymmetric gauge theories possess Wilson--'t Hooft lines
which preserve half of the eight supercharges and carry both electric
and magnetic charges~\cite{Kapustin:2005py}.  The line defects that
appear on the gauge theory side of the correspondence are such dyonic
Wilson--'t Hooft lines.  Roughly speaking, the statement of the
correspondence is that the vacuum expectation values (vevs) of certain
Wilson--'t Hooft lines are equal to classical values of transfer
matrices of certain quantum integrable systems.

A more precise statement is as follows. Consider the theory described
by an $n$-node circular quiver whose gauge group is the product of $n$
copies of $\SU(N)$.  Let $m^1$, $\dotsc$, $m^n$ be the mass parameters
of the $n$ bifundamental hypermultiplets.  We place the theory on
$S^1 \times_\eps \R^2 \times \R$ and wind a Wilson--'t Hooft line
$T_{\square,\sigma}$ around $S^1$, where $\eps$ is a twist parameter.
The line operator $T_{\square,\sigma}$ is magnetically charged
uniformly under the $\SU(N)$ factors; it transforms in the vector
representation of the Langlands dual of each $\SU(N)$.  The electric
charge is specified by an $n$-tuple of signs
$\sigma = (\sigma^1, \dotsc, \sigma^n)$.  On the Coulomb branch of
vacua, the vev $\vev{T_{\square,\sigma}}$ of $T_{\square,\sigma}$ can
be expressed in terms of trigonometric functions of the vevs of some
vector multiplet fields.  With respect to an appropriate holomorphic
symplectic structure, the Weyl quantization of
$\vev{T_{\square,\sigma}}$ gives a transfer matrix $\CT_{\sigma,m}$ of
a quantum integrable system, constructed from $n$ L-operators
$\CL_{\sigma^1,m^1}$, $\dotsc$, $\CL_{\sigma^n,m^n}$.

In section~\ref{sec:QIS}, we introduce the L-operators $\CL_{\pm,m}$,
which are the basic ingredients of the integrable system side, as
particular trigonometric limits of the elliptic L-operator found
in~\cite{MR1463830}.  The elliptic L-operator satisfies the RLL
relation with the elliptic dynamical R-matrix~\cite{Baxter:1972wf,
  MR908997, Jimbo:1987mu}.  Consequently, its transfer matrix defines
a quantum integrable system according to the standard procedure, as we
review in this section.  The key result of this section is the
expression~\eqref{eq:T-trig} for the Wigner transform
$\vev{\CT_{\sigma,m}}$ of the transfer matrix $\CT_{\sigma,m}$, which
is the inverse of the Weyl quantization
$\vev{\CT_{\sigma,m}} \mapsto \CT_{\sigma,m}$.

We will establish the correspondence by supersymmetric
localization~\cite{Pestun:2016zxk}, a technique to exactly compute the
path integral for supersymmetric observables in theories with
sufficiently large supersymmetry, placed on specific spacetime
geometries.  In the present setup, the relevant computation was
carried out by Ito, Okuda and Taki~\cite{Ito:2011ea}.

In section~\ref{sec:gauge}, we apply the formula obtained
in~\cite{Ito:2011ea} to the Wilson--'t Hooft line $T_{\square,\sigma}$
and show that its vev reproduces the Wigner transform of the transfer
matrix $\CT_{\sigma,m}$:
\begin{equation}
  \vev{T_{\square,\sigma}} = \vev{\CT_{\sigma,m}} \,.
\end{equation}
This is the main result of the paper.  We will also explain how to
recover L-operators and monodromy matrices from theories described by
linear quivers.  Moreover, we will propose a generalization of the
correspondence to a broader class of Wilson--'t Hooft lines, in which
the vector representation $\square$ is replaced by other
representations.

Another route to compute the Wilson--'t Hooft line vev is via the AGT
correspondence~\cite{Alday:2009aq, Wyllard:2009hg}.  For a large class
of $\CN = 2$ supersymmetric field theories, which includes the
circular quiver theory, the AGT correspondence states that the
partition function on an ellipsoid is equal to a correlation function
in a two-dimensional conformal field theory (CFT), namely Toda theory.
Under this correspondence, Wilson--'t Hooft lines in the former are
mapped to line defects in the latter~\cite{Alday:2009fs,
  Drukker:2009id, Gomis:2010kv}.  These line defects are known as
Verlinde operators~\cite{Verlinde:1988sn}.

In section~\ref{sec:Toda}, we give an alternative derivation of the
correspondence between Wilson--'t Hooft lines and transfer matrices
that utilizes a variant of the AGT correspondence, proposed
in~\cite{Ito:2011ea}, relating Verlinde operators to Wilson--'t Hooft
lines in $S^1 \times_\eps \R^2 \times \R$, rather than in an
ellipsoid.  We consider the Verlinde operator corresponding to the
Wilson--'t Hooft line in question, and verify that its action on
conformal blocks matches the action of the transfer matrix in the
quantum integrable system.

While the correspondence between Wilson--'t Hooft lines and transfer
matrices can be established by comparison of concrete calculations,
mere matching does not explain why the correspondence exists in the
first place.

In section~\ref{sec:branes}, we provide an explanation using string
theory.  We will realize the circular quiver theory and the Wilson--'t
Hooft line by branes, and apply string dualities to map the brane
configuration to another one that realizes Costello's four-dimensional
Chern--Simons theory~\cite{Costello:2013zra, Costello:2017dso} and
line defects in it.  Four-dimensional Chern--Simons theory depends
topologically on two directions, which form a cylinder in our case,
and holomorphically on the remaining two directions.  From this
property it follows that line defects extending in the periodic
topological direction produce transfer matrices of quantum integrable
systems.

The embedding into string theory puts the correspondence treated in
this paper in a bigger context.  As discussed
in~\cite{Costello:2018txb}, by brane realization and dualities,
four-dimensional Chern--Simons theory is related to other field theory
setups in which the same kind of integrability was found to arise.  In
particular, there is a duality frame that realizes the trigonometric
limit of the setup of~\cite{Maruyoshi:2016caf, Yagi:2017hmj}, thus
connecting to the correspondence mentioned at the beginning.

\section{Transfer matrices for Wilson--'t Hooft lines}
\label{sec:QIS}

In this section we discuss the integrable system side of the
correspondence.  After reviewing L-operators, transfer matrices and
their relation to quantum integrable systems, we introduce an
L-operator for the elliptic dynamical R-matrix.  Then we define
fundamental trigonometric L-operators as certain limits of the
elliptic L-operator.  These fundamental L-operators are building
blocks of transfer matrices that correspond to Wilson--'t Hooft lines
in $\CN = 2$ supersymmetric circular quiver theories.

\subsection{L-operators and quantum integrable systems}

Let $\hf$ be a finite-dimensional commutative complex Lie algebra and
$V$ a finite-dimensional diagonalizable $\hf$-module.  Choosing a
basis $\{v_i\}$ of $V$ that is homogeneous with respect to weight
decomposition, we denote the weight of $v_i$ by $h_i$ and the
$(i,j)$th entry of a matrix $M \in \End(V)$ by $M^i_j$.  We write
$\CM_{\hf^*}$ for the field of meromorphic functions on the dual
space $\hf^*$ of $\hf$.

Let $R\colon \C \times \hf^* \to \End(V \otimes V)$ be an
$\End(V \otimes V)$-valued meromorphic function on
$\C \times \hf^*$ that is invertible at a generic point
$(z,a) \in \C \times \hf^*$.  The coordinate $z$ is called
the \emph{spectral parameter} and $a$ is called the
\emph{dynamical parameter}.

In the discussions that follow, fundamental roles will be played by
L-operators.  By an \emph{L-operator} for $R$, we mean a map
$L\colon \C \to \End(V \otimes \CM_{\hf^*} \otimes
\CM_{\hf^*})$, which we think of as a matrix whose entries are
linear operators on meromorphic functions on
$\hf^* \times \hf^*$.%
\footnote{Our definition of L-operators is more general than the one
  given in~\cite{MR1645196} in that L-operators may depend on two
  independent dynamical parameters.  This generalization is necessary
  in order to treat the elliptic L-operator appearing
  in~\cite{MR1463830} in the formalism of dynamical R-matrix.}
It must satisfy two conditions.

First, its matrix elements act on
$f \in \CM_{\hf^*} \otimes \CM_{\hf^*}$ as
\begin{equation}
  L(z)^j_i f(a^1,a^2)
  = L(z;a^1,a^2)^j_i \Delta_i^1 \Delta_j^2 f(a^1,a^2) \,,
\end{equation}
where $L(z;a^1,a^2)^j_i$ is a meromorphic function on
$\C \times \hf^* \times \hf^*$ and $\Delta_i^1$,
$\Delta_j^2$ are difference operators such that
\begin{equation}
  \Delta_i^1 f(a^1,a^2)
  = f(a^1 - \eps h_i,a^2) \,,
  \qquad
  \Delta_j^2 f(a^1,a^2)
  = f(a^1, a^2 - \eps h_j) \,.
\end{equation}
Here $\eps$ is a fixed complex parameter.

Second, the L-operator satisfies the \emph{RLL relation}
\begin{multline}
  \label{eq:RLL}
  \sum_{k,l}
  R(z-z', a^2)^{mn}_{kl}
  L(z;a^1,a^2)^k_i
  L(z'; a^1 - \eps h_i, a^2 - \eps h_k)^l_j
  \\
  =
  \sum_{k,l}
  L(z';a^1,a^2)^n_l
  L(z; a^1 - \eps h_l, a^2 - \eps h_n)^m_k
  R(z-z', a^1)^{kl}_{ij} \,.
\end{multline}
Equivalently, the operator relation
\begin{equation}
  \sum_{k,l}
  R(z-z', a^2)^{mn}_{kl} L(z)^k_i L(z')^l_j
  =
  \sum_{k,l}
  R(z-z', a^1)^{kl}_{ij} L(z')^n_l L(z)^m_k
\end{equation}
holds on any meromorphic function $f(a^1, a^2)$.

It is helpful, and will turn out to be physically meaningful, to
represent the L-operator graphically as two crossing oriented line
segments:
\begin{equation}
  L(z)
  =
  \begin{tikzpicture}[xscale=0.8, yscale=0.5, baseline=(x.base)]
    \node (x) at (0,0) {\vphantom{x}};

    \draw[thick, ->] (0,0) node[left] {$z$} -- (2,0);
    \draw[thick, double, ->] (1,-1) -- (1,1);
  \end{tikzpicture}
  \quad .
\end{equation}
The solid line extending in the horizontal direction has a spectral
parameter.  The graphical representation of a matrix element of the
L-operator is
\begin{equation}
  L(z; a^1, a^2)_i^j
  =
  \begin{tikzpicture}[xscale=1.5, baseline=(x.base)]
    \node (x) at (0,0) {\vphantom{x}};

    \draw[thick, ->] (0,0) node[left] {$z$} -- (2,0);
    \draw[thick, double, ->] (1,-1) -- (1,1);

    \node at (0.5, 0.5) {$a^1$};
    \node at (1.5, 0.5) {$a^2$};
    \node at (0.5, -0.5) {$a^1 - \eps h_i$};
    \node at (1.5, -0.5) {$a^2 - \eps h_j$};
    \node[spin] at (0.5,0) {$i$};
    \node[spin] at (1.5,0) {$j$};
  \end{tikzpicture}
  \quad .
\end{equation}
Each edge of a solid line carries a state in $V$, and the state may
change when the line crosses another line.  To each region separated
by lines, a dynamical parameter is assigned.  The values of dynamical
parameters on the two sides of a solid line carrying state $v_i$
differ by $\eps h_i$.

We also represent the operator $R$ as two crossing solid lines:
\begin{equation}
  R(z - z', a)_{ij}^{kl}
  =
  \begin{tikzpicture}[scale=1, baseline=(x.base)]
    \node (x) at (0,0) {\vphantom{x}};

    \draw[thick, ->] (0,0) node[left] {$z$} -- (2,0);
    \draw[thick, ->] (1,-1) node[below] {$z'$} -- (1,1);

    \node at (0.5, 0.5) {$a$};

    \node[spin] at (0.5,0) {$i$};
    \node[spin] at (1.5,0) {$k$};
    \node[spin] at (1,-0.5) {$j$};
    \node[spin] at (1,0.5) {$l$};
  \end{tikzpicture}
  \quad .
\end{equation}
Then, the RLL relation~\eqref{eq:RLL} simply means an equality between
two configurations involving two solid and one double lines:
\begin{equation}
    \begin{tikzpicture}[yscale=0.6, baseline=(x.base)]
      \node (x) at (30:2) {};
      
      \draw[thick, ->] (0,0) node[left] {$z'$} -- ++(30:3);
      \draw[thick, ->] (0,2) node[left] {$z$} -- ++(-30:3);
      \draw[thick, double, ->] (-30:1) -- ++(0,3);
      
      \node (O) at ({sqrt(3)*4/6},1) {};
      \node[yshift=2pt] at ($(O) + (120:{sqrt(3)*4/6})$) {$a^1$};
      \node at ($(O) + (60:{sqrt(3)*3/6})$) {$a^2$};
    \end{tikzpicture}
    \ =
    \begin{tikzpicture}[yscale=0.6, baseline=(x.base)]
      \node (x) at (30:1) {};
      
      \draw[thick, ->] (0,0) node[left] {$z'$} -- ++(30:3);
      \draw[thick, ->] (0,1) node[left] {$z$} -- ++(-30:3);
      \draw[thick, double, ->] (-30:2) -- ++(0,3);
      
      \node (O) at ({sqrt(3)*5/6},0.5) {};
      \node at ($(O) + (120:{sqrt(3)*3/6})$) {$a^1$};
      \node[yshift=2pt, xshift=2pt]  at ($(O) + (60:{sqrt(3)*4/6})$) {$a^2$};
    \end{tikzpicture}
    \quad .
\end{equation}
The states carried by the internal solid edges are summed over.

By comparing the values of the dynamical parameter assigned to the
lower right regions of the two sides, we see that for $R$ to satisfy
the RLL relation with some L-operators, generally it must commute with
$h \otimes 1 + 1 \otimes h$ for all $h \in \hf$; in other words,
$R(z,a)_{ij}^{kl} = 0$ unless $h_i + h_j = h_k + h_l$.  This is a
consistency condition for the rule that determines how dynamical
parameters change across solid lines.

Associated with an L-operator, there is an integrable quantum
mechanical system consisting of particles moving in the space $\hf^*$.
The Hilbert space of each particle is $\CM_{\hf^*}$.  (This is quantum
mechanics in which real variables are analytically continued to
complex ones.) The Hilbert space of the system is
$\CM_{\hf^*}^{\otimes n}$ if $n$ is the number of particles.

To construct this system, define the \emph{monodromy matrix}
$M\colon \C \to \End(V \otimes \CM_{\hf^*}^{\otimes n + 1})$ by the
product of $n$ copies of the L-operator: its matrix elements are given
by
\begin{equation}
  M(z)^{i^{n+1}}_{i^1}
  =
  \sum_{i^2, \dotsc, i^n}
  \prod_{r=1}^n
  L(z; a^r,a^{r+1})^{i^{r+1}}_{i^r}
  \prod_{s=1}^{n+1} \Delta_{i^s}^s \,,
\end{equation}
acting on any meromorphic function $f(a^1, \dotsc, a^{n+1})$.  (The
superscript on $\Delta_i$ specifies the variable on which the
difference operator acts.)  This is a solid line crossing $n$ double
lines:
\begin{equation}
  M(z)
  =
  \begin{tikzpicture}[xscale=0.8, yscale=0.5, baseline=(x.base)]
    \node (x) at (0,0) {\vphantom{x}};

    \draw[thick, ->] (0,0) node[left] {$z$} -- (5,0);
    \draw[thick, double, ->] (1,-1) -- +(90:2);
    \draw[thick, double, ->] (2,-1) -- +(90:2);
    
    \draw[ultra thick, dotted, -] (2.55,-0.5) -- (3.45,-0.5);

    \draw[thick, double, ->] (4,-1) -- +(90:2);
  \end{tikzpicture}
  \quad .
\end{equation}
Identifying $a^{n+1} = a^1$ and taking the trace, one
obtains the \emph{transfer matrix}
$T\colon \C \to \End(\CM_{\hf^*}^{\otimes n})$:
\begin{equation}
  T(z)
  =
  \sum_{i^1, \dotsc, i^n}
  \prod_{r=1}^n
  L(z; a^r,a^{r+1})^{i^{r+1}}_{i^r}
  \prod_{s=1}^n \Delta_{i^s}^s
  \,,
  \qquad
  i^{n+1} = i^1 \,.
\end{equation}
Graphically, $T(z)$ is represented by the same picture as above but
with the horizontal direction made periodic.

By construction, $T$ is an $\End(\CM_{\hf^*}^{\otimes n})$-valued
meromorphic function.  As such, each coefficient $T_m$ in the Laurent
expansion $T(z) = \sum_{m \in \Z} T_m z^m$ is an operator acting on
the Hilbert space $\CM_{\hf^*}^{\otimes n}$.  Then, one may pick a
particular linear combination of these coefficients and declare that
it is the Hamiltonian of the quantum mechanical system.  The
Hamiltonian thus obtained is a difference operator, which is typical
of relativistic systems.

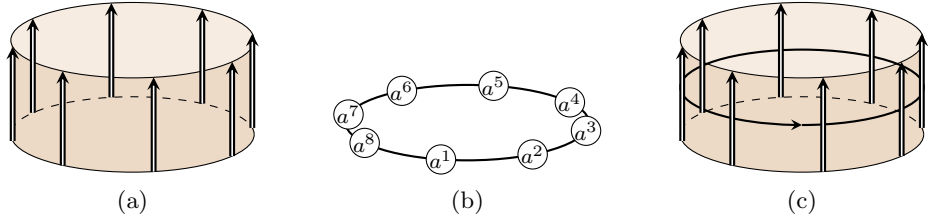
\begin{figure}
  \centering
  \subfloat[\label{fig:spin-chain-no-T}]{
    \begin{tikzpicture}[yscale=0.25, xscale=0.8]
      \fill[brown!30] (-2,0) arc (-180:0:2) -- +(0,5) arc (0:-180:2) --
      +(0,-5) -- cycle;
      \fill[brown!15]  (0,5) circle [radius=2];      

      \draw[dashed] (2,0) arc (0:180:2);
      \draw (-2,0) arc (180:360:2);
      \draw (0,5) circle [radius=2];
      
      \draw[thick, double, ->] (10:2) -- ++(90:5);
      \draw[thick, double, ->] (55:2) -- ++(90:5);
      \draw[thick, double, ->] (100:2) -- ++(90:5);
      \draw[thick, double, ->] (145:2) -- ++(90:5);
      \draw[thick, double, ->] (190:2) -- ++(90:5);
      \draw[thick, double, ->] (235:2) -- ++(90:5);
      \draw[thick, double, ->] (280:2) -- ++(90:5);
      \draw[thick, double, ->] (325:2) -- ++(90:5);
    \end{tikzpicture}
  }
  \qquad
  \subfloat[\label{fig:spin-chain-chain}]{
    \begin{tikzpicture}[yscale=0.25, xscale=0.8]
      \draw[thick] (0,0) circle [radius=2];

      \node[spin, font=\scriptsize] at ({10-45/2}:2) {$a^3$};
      \node[spin, font=\scriptsize] at ({55-45/2}:2) {$a^4$};
      \node[spin, font=\scriptsize] at ({100-45/2}:2) {$a^5$};
      \node[spin, font=\scriptsize] at ({145-45/2}:2) {$a^6$};
      \node[spin, font=\scriptsize] at ({190-45/2}:2) {$a^7$};
      \node[spin, font=\scriptsize] at ({235-45/2}:2) {$a^8$};
      \node[spin, font=\scriptsize] at ({280-45/2}:2) {$a^1$};
      \node[spin, font=\scriptsize] at ({325-45/2}:2) {$a^2$};
    \end{tikzpicture}
  }
  \qquad
  \subfloat[\label{fig:spin-chain-T}]{
    \begin{tikzpicture}[yscale=0.25, xscale=0.8]
      \fill[brown!30] (-2,0) arc (-180:0:2) -- +(0,5) arc (0:-180:2) --
      +(0,-5) -- cycle;
      \fill[brown!15]  (0,5) circle [radius=2];      

      \draw[dashed] (2,0) arc (0:180:2);
      \draw (-2,0) arc (180:360:2);
      \draw (0,5) circle [radius=2];
      
      \draw[thick, ->] (-2,2.5) arc (-180:-90:2);
      \draw[thick] (-2,2.5) arc (180:-90:2);
      
      \draw[thick, double, ->] (10:2) -- ++(90:5);
      \draw[thick, double, ->] (55:2) -- ++(90:5);
      \draw[thick, double, ->] (100:2) -- ++(90:5);
      \draw[thick, double, ->] (145:2) -- ++(90:5);
      \draw[thick, double, ->] (190:2) -- ++(90:5);
      \draw[thick, double, ->] (235:2) -- ++(90:5);
      \draw[thick, double, ->] (280:2) -- ++(90:5);
      \draw[thick, double, ->] (325:2) -- ++(90:5);
    \end{tikzpicture}
  }
  \caption{(a) Double lines in the longitudinal direction of a
    cylinder.  (b) The corresponding quantum spin chain with
    continuous spin variables.  (c) A solid line winding around the
    cylinder acts on the spin chain by the transfer matrix.}
  \label{fig:spin-chain}
\end{figure}

Alternatively, one may think of this system as a one-dimensional
periodic quantum spin chain.  This spin chain is constructed from $n$
double lines extending in the longitudinal direction of a cylinder, as
shown in figure~\ref{fig:spin-chain-no-T}.  The dynamical parameter
$a^r$ resides in the region sandwiched by the $r$th and the
$(r+1)$th double lines.  One regards the $n$ dynamical parameters
$a^1$, $\dotsc$, $a^n$ as continuous spin variables; see
figure~\ref{fig:spin-chain-chain}.  Thinking of the longitudinal
direction as the time direction, the Hilbert space of the spin chain
is again $\CM_{\hf^*}^{\otimes n}$.  An action of $T(z)$ on the
Hilbert space is induced by an insertion of a solid line with spectral
parameter $z$ in the circumferential direction of the cylinder, as in
figure~\ref{fig:spin-chain-T}.

The integrability of the system is a consequence of the RLL relation.
By repeated use of the RLL relation, one deduces that the monodromy
matrix satisfies a similar relation:
\begin{equation}
  \sum_{k,l}
  R(z-z', a^{n+1})^{mn}_{kl}
  M(z)^k_i M(z')^l_j
  =
  \sum_{k,l}
  R(z-z', a^1)^{kl}_{ij}
  M(z')^n_l M(z)^m_k
  \,.
\end{equation}
Multiplying both sides by $R^{-1}(z-z', a^1)^{ij}_{mn}$, setting
$a^{n+1} = a^1$ and summing over $i$, $j$, $m$, $n$, one
finds
\begin{equation}
  T(z) T(z') = T(z') T(z) \,.
\end{equation}
In other words, transfer matrices at different values of the spectral
parameter commute.  It follows that the Laurent coefficients $\{T_m\}$
mutually commute and, in particular, commute with the Hamiltonian.
Hence, the system has a series of commuting conserved charges.

There is a slight generalization of the above construction of
commuting transfer matrices.  Suppose that $g \in \End(V)$ satisfies
\begin{equation}
  (g \otimes g) R(z,a)
  =  R(z,a) (g \otimes g)
\end{equation}
and a subspace $W$ of $V$ is invariant under $R$, $R^{-1}$, $L$ and
$g$.  (For instance, the invariance of $W$ under $R$ means that
$R(z,a)(W \otimes V) \subset W \otimes V$ and
$R(z,a)(V \otimes W) \subset V \otimes W$ for all $z$, $a$.)  Then,
the trace can be twisted by $g$ and restricted to $W$:
\begin{equation}
  T_{g,W} = \Tr_W(gM) \,.
\end{equation}
If $W_1$, $W_2$ are such invariant subspaces, then
\begin{equation}
  [T_{g,W_1}(z), T_{g,W_2}(z')] = 0 \,.
\end{equation}
Thus, we get different kinds of transfer matrices labeled by invariant
subspaces, and they commute with each other.  A typical situation in
which this construction applies is when $\hf$ is a Cartan subalgebra
of a complex Lie algebra $\gf_\C$, $V$ is a direct sum of irreducible
representations of $\gf_\C$, and $g$ is an element of $\gf_\C$.

Algebraically, L-operators give representations of \emph{dynamical
  quantum groups}~\cite{Felder:1994be, Felder:1994pb, MR1645196}.  As
an algebra, the dynamical quantum group corresponding to $R$ is
generated by the meromorphic functions on
$\C \times \hf^* \times \hf^*$, together with additional generators
$l(z)^i_j$, $l^{-1}(z)^i_j$.  The generators $l(z)^i_j$ are to be
understood as the matrix elements of an abstract L-operator and
satisfy the same relations as above; $l^{-1}(z)^i_j$ are the elements
of the inverse matrix.  This algebra has further structures (coproduct
and counit) which make it an $\hf$-bialgebroid.

\subsection{Elliptic L-operator}

An important example of an L-operator is one for the elliptic
dynamical R-matrix~\cite{Baxter:1972wf, MR908997, Jimbo:1987mu}, which
is a representation of the elliptic quantum group for $\slf_N$.  In
this example, $\hf$ is the Cartan subalgebra of $\slf_N$ and
$V = \C^N$ is the vector representation of $\slf_N$.

The Lie algebra $\slf_N$ consists of the traceless complex
$N \times N$ matrices and $\hf$ is the subalgebra of diagonal
elements.  We denote by $E_{ij} \in \glf_N$ the matrix that has $1$ in
the $(i,j)$th entry and $0$ elsewhere, and by $E^*_{ij}$ the element
of $\glf_N^* = \Hom(\glf_N, \C)$ such that
$\langle E_{ij}, E_{kl}^*\rangle = \delta_{ik} \delta_{jl}$.  (The
bilinear map $\langle-,-\rangle\colon \glf_N \times \glf_N^* \to \C$
is the natural pairing.)  The elements of $\hf$ are matrices of the
form $\sum_{i=1}^N b_i E_{ii}$, with $\sum_{i=1}^N b_i = 0$.  Since
$\hf$ is isomorphic to the quotient of the subspace of $\glf_N$
consisting of the diagonal matrices by the subspace spanned by the
identity matrix $I = \sum_{i=1}^N E_{ii}$, the dual space $\hf^*$ is
isomorphic to the subspace of $\glf_N^*$ consisting of elements of the
form $\sum_{i=1}^N a_i E_{ii}^*$ such that
$\langle I, \sum_{i=1}^N a_i E_{ii}^*\rangle = \sum_{i=1}^N a_i = 0$.
Thus, $\hf^*$ may also be identified with the space of traceless
diagonal matrices.

The natural action of $\slf_N$ on $\C^N$ defines the vector
representation of $\slf_N$.  In terms of the standard basis
$\{e_1, \dotsc, e_N\}$ of $\C^N$, we have
$\sum_{j=1}^N a_j E_{jj} e_i = a_i e_i$.  The weight of $e_i$ is
therefore
\begin{equation}
  h_i
  = E_{ii}^* - \frac{1}{N} \sum_{j=1}^N E_{jj}^* \,.
\end{equation}
For $a \in \hf^*$, we write $a_i = \langle E_{ii}, a\rangle$.  Then,
$\sum_{i=1}^N a_i = 0$ and
$a = \sum_{i=1}^N a_i E_{ii}^* = \sum_{i=1}^N a_i h_i$.

Fix a point $\tau$ in the upper half plane, $\Im\tau > 0$, and let
 \begin{equation}
  \theta_1(z)
  =
  - \sum_{j \in \Z + \frac12} e^{\pi\iu j^2\tau + 2\pi\iu j(z + \frac12)}
\end{equation}
be Jacobi's first theta function.  The \emph{elliptic dynamical
  R-matrix} $R^{\text{ell}}$ is defined by~\cite{Felder:1994be,
  Felder:1994pb, MR1645196}
\begin{equation}
  R^{\text{ell}}(z,a)
  =
  \sum_{i=1}^N E_{ii} \otimes E_{ii}
  + \sum_{i \neq j} \alpha(z,a_{ij}) E_{ii} \otimes E_{jj}
  + \sum_{i \neq j} \beta(z,a_{ij}) E_{ji} \otimes E_{ij}
  \,,
\end{equation}
where $a_{ij} = a_i - a_j$ and
\begin{equation}
  \alpha(z,a)
  = \frac{\theta_1(a+\eps) \theta_1(-z)}
         {\theta_1(a) \theta_1(\eps - z)} \,,
  \qquad
  \beta(z,a)
  = \frac{\theta_1(a - z) \theta_1(\eps)}
          {\theta_1(a) \theta_1(\eps - z)} \,.
\end{equation}

The \emph{elliptic L-operator} $L^{\text{ell}}$, which satisfies the
RLL relation with $R^{\text{ell}}$, has the matrix elements given
by~\cite{MR1463830}
\begin{equation}
  \label{eq:L-ell}
  L^{\text{ell}}_{w,y}(z; a^1, a^2)^j_i
  =
  \frac{\theta_1(z - w + a^2_j - a^1_i)}{\theta_1(z - w)}
  \prod_{k (\neq i)}
  \frac{\theta_1(a^1_k - a^2_j - y)}{\theta_1(a^1_{ki})}
  \,.
\end{equation}
The complex numbers $w$, $y$ may be thought of as spectral parameters
for the corresponding double line.  The presence of the two parameters
is due to the fact that $R^{\text{ell}}(z,a)$ is invariant under shift
of $a$ by a multiple of the identity matrix $I$ and in the RLL
relation~\eqref{eq:RLL} the spectral parameters $z$, $z'$ enter the
R-matrix only through the difference $z - z'$; note also that the
L-operator can be multiplied by any function of the spectral
parameter.

The elliptic dynamical R-matrix and the elliptic L-operator have many
more properties than just that they satisfy the RLL relation.  Most
importantly, the R-matrix is a solution of the dynamical Yang--Baxter
equation~\cite{Gervais:1983ry, Felder:1994be, Felder:1994pb} and
encodes the Boltzmann weights for a two-dimensional integrable lattice
model~\cite{Baxter:1972wf, MR908997, Jimbo:1987mu}.  This model is
equivalent to the eight-vertex model~\cite{Baxter:1971cr,
  Baxter:1972hz} (or more precisely, the Belavin
model~\cite{Belavin:1981ix} which is an $\slf_N$ generalization of the
eight-vertex model) in the sense that the transfer matrices of the two
models are related by a similarity transformation.  The elliptic
L-operator, on the other hand, satisfies the RLL relation with another
R-matrix which describes an integrable lattice model called the
Bazhanov--Sergeev model~\cite{Bazhanov:2010kz, Bazhanov:2011mz}, whose
spins variables take values in $\hf^*$.  We will not discuss these
aspects in this paper.  The interested reader is referred to
\cite{Yagi:2017hmj} for more details.

\subsection{Trigonometric L-operators}

The L-operators that appear in the correspondence with Wilson--'t
Hooft lines are obtained from the elliptic L-operator $L^{\text{ell}}$
via the trigonometric limit $\tau \to \iu\infty$.  For comparison with
gauge theory results, we actually need to express these L-operators in
somewhat different forms.

First, we describe L-operators in a quantum mechanical language.  Let
us explain this description in the case in which $\hf$ is the Cartan
subalgebra of $\slf_N$.  Recall that $\slf_N$ has simple coroots
\begin{equation}
  \alpha^\vee_i = E_{ii} - E_{i+1,i+1} \,,
  \qquad
  i = 1, \, \dotsc, \, N-1 \,,
\end{equation}
and the fundamental weights
\begin{equation}
  \omega_i = (\alpha^\vee_i)^* = \sum_{j=1}^i h_j \,.
\end{equation}

Consider quantum mechanics of a particle living in
$\hf^* \times \hf^*$, with Planck constant
\begin{equation}
  \hbar = -\frac{\eps}{2\pi} \,.
\end{equation}
If $(a^1,a^2) \in \hf^* \times \hf^*$ is the position of the particle,
we write $a^r = \sum_{i=1}^{N-1} q^r_i \omega_i$, $r = 1$, $2$.
Similarly, we write the momenta $(b^1, b^2) \in \hf \times \hf$ of the
particle as $b^r = \sum_{i=1}^{N-1} p^r_i \alpha^\vee_i$.  The
corresponding position and momentum operators $\qh^r_i$, $\ph^s_i$
satisfy the canonical commutation relations:
\begin{equation}
  [\qh_i^r, \ph_j^s] = \iu\hbar \delta^{rs} \delta_{ij} \,,
  \qquad
  i, \, j = 1, \, \dotsc, \, N-1 \,.
\end{equation}
(As before, we are treating $q^r_i$, $p^r_i$ as analytically continued
variables.)

To rewrite the commutation relations in a form that is invariant under
the action of the Weyl group, we make a change of basis
\begin{equation}
  a^r = \sum_{i=1}^N a^r_i E_{ii}^* \,,
  \qquad
  b^r = \sum_{i=1}^N b^r_i E_{ii} \,.
\end{equation}
Then, the corresponding observables $\ah^r_i$, $\bh^r_i$ obey the
traceless condition,
$\sum_{i=1}^N \ah^r_i = \sum_{i=1}^N \bh^r_i = 0$, and satisfy the
commutation relations
\begin{equation}
  [\ah_i^r, \bh_j^s]
  = \iu\hbar \delta^{rs} \Bigl(\delta_{ij} - \frac{1}{N}\Bigr) \,,
  \qquad
  i, \, j = 1, \, \dotsc, \, N \,.
\end{equation}
Using these observables we can identify the matrix elements of an
L-operator $L$ with an operator in the Hilbert space of this quantum
mechanical system:
\begin{equation}
  \label{eq:L-in-QM}
  L(z)_i^j
  = L(z; \ah^1, \ah^2)_i^j e^{2\pi\iu(\bh_i^1 + \bh_j^2)} \,.
\end{equation}

In quantum mechanics, there is an invertible map from functions on the
classical phase space to operators in the Hilbert space, known as the
Weyl transform: if $q$ and $p$ are canonically conjugate variables, it
maps
\begin{equation}
  f(q,p)
  \mapsto
  \fh(\qh,\ph)
  =
  \int_{\R^4}
  \rmd x \, \rmd y \, \rmd p \, \rmd q \, 
  f(q,p)
  e^{\iu(x(\qh - q) + y(\ph - p))} \,.
\end{equation}
The inverse map is the \emph{Wigner transform}, which we denote by
$\vev{-}$:
\begin{equation}
  f(\qh,\ph)
  \mapsto
  \vev{f(\qh,\ph)}
  = \int_\R \rmd x \, e^{\iu px/\hbar}
  \Bigbra{q + \frac12 x} \fh(\qh,\ph) \Bigket{q - \frac12 x} \,.
\end{equation}
In the situation at hand, if we rewrite the
expression~\eqref{eq:L-in-QM} as
\begin{equation}
  L(z)_i^j
  =
  e^{\pi\iu(\bh_i^1 + \bh_j^2)}
  \Lt(z; \ah^1, \ah^2)_i^j
  e^{\pi\iu(\bh_i^1 + \bh_j^2)}
  \,,
\end{equation}
then we have
\begin{equation}
  \vev{L(z)_i^j}
  =
  e^{2\pi\iu(b_i^1 + b_j^2)} \Lt(z; a^1, a^2)_i^j \,.
\end{equation}

Next, we apply a similarity transformation to the elliptic L-operator.
Assume $\Im\eps > 0$ and let
\begin{equation}
  \Gamma(z,\tau,\eps)
  =
  \prod_{m,n=0}^\infty
  \frac{1 - e^{2\pi\iu((m+1)\tau + (n+1)\eps - z)}}
       {1 - e^{2\pi\iu(m\tau + n\eps + z)}}
\end{equation}
be the elliptic gamma function.  Then,
$\Gammab(z) = e^{\pi\iu z^2/2\eps} \Gamma(z,\tau,\eps)$ has the
property that
$\Gammab(z + \eps, \tau,\eps) = g(\tau,\eps) \theta_1(z) \Gammab(z,
\tau,\eps)$ for some function $g(\tau,\eps)$.  We define the
conjugated L-operator $\CL^{\text{ell}}_{w,m}(z)$ by
\begin{equation}
  \CL^{\text{ell}}_{w,m}(z)_i^j
  =
  \Phi_{m-\frac12\eps}
  L^{\text{ell}}_{w,m - \frac12\eps}(z)_i^j
  \Phi_{m-\frac12\eps}^{-1} \,,
\end{equation}
where
\begin{equation}
  \Phi_y
  =
  \prod_{k,l=1}^N \Gammab(\ah^1_k - \ah^2_l - y)^{\frac12}
  \prod_{k \neq l} \Gammab(\ah^1_{kl})^{-\frac12} \,.
\end{equation}
It has the Wigner transform
\begin{multline}
  \vev{\CL^{\text{ell}}_{w,m}(z)_i^j}
  =
  e^{2\pi\iu(b_i^1 + b_j^2)}
  \frac{\theta_1(z - w + a^2_j - a^1_i)}{\theta_1(z - w)}
  \\
  \times
  \Biggl(
  \frac{\prod_{k (\neq i)} \theta_1(a^1_k - a^2_j - m)
        \prod_{l (\neq j)} \theta_1(a^1_i - a^2_l - m)}
       {\prod_{k (\neq i)}\theta_1(a^1_{ki} - \frac12\eps)
        \theta_1(a^1_{ik} - \frac12\eps)}
  \Biggr)^{\frac12}
  \,.
\end{multline}

With these preparations, let us finally take the trigonometric limit
to define the trigonometric L-operator:
\begin{equation}
  \CL_{w,m}
  = \lim_{\tau \to \iu\infty} \CL^{\text{ell}}_{w,m} \,.
\end{equation}
The trigonometric L-operator satisfies the RLL relation with the
trigonometric limit $R^{\text{trig}}$. of the elliptic R-matrix
$R^{\text{ell}}$.  Concretely, $\CL_{w,m}$ and $R^{\text{trig}}$ are
obtained from $\CL^{\text{ell}}_{w,m}$ and $R^{\text{ell}}$ by the
replacement $\theta_1(z) \to \sin(\pi z)$.

Once we are in the trigonometric setup, the quasi-periodicity in
$z \to z + \tau$ is lost and we can further take the limits
$w \to \pm\iu\infty$.  This allows us to introduce more fundamental
L-operators:
\begin{equation}
  \label{eq:fund-L}
  \CL_{\pm,m} = \lim_{w \to \pm\iu\infty} \CL_{w,m} \,.
\end{equation}
These L-operators do not depend on the spectral parameters $z$, $w$,
and their matrix elements have the Wigner transforms
\begin{equation}
  \label{eq:fund-L-ij}
  \vev{(\CL_{\pm,m})_i^j}
  =
  e^{2\pi\iu(b_i^1 + b_j^2)}
  e^{\pm\pi\iu(a^2_j - a^1_i)}
  \ell_m(a^1, a^2)_i^j \,,
\end{equation}
with
\begin{equation}
  \label{eq:ell}
  \ell_m(a^1, a^2)_i^j
  =
  \Biggl(
  \frac{\prod_{k (\neq i)} \sin\pi(a^1_k - a^2_j - m)
        \prod_{l (\neq j)} \sin\pi(a^1_i - a^2_l - m)}
       {\prod_{k (\neq i)}\sin\pi(a^1_{ki} - \frac12\eps)
        \sin\pi(a^1_{ik} - \frac12\eps)}
  \Biggr)^{\frac12}
  \,.
\end{equation}
The L-operator for arbitrary parameters $z$, $w$ can be realized as a
linear combination of $\CL_{\pm,m}$:
\begin{equation}
  \CL_{w,m}(z)
  =
  \frac{e^{\pi\iu(z-w)} \CL_{+,m} - e^{-\pi\iu(z-w)} \CL_{-,m}}{\sin\pi(z-w)}
  \,.
\end{equation}

The monodromy matrix $\CM_{\sigma,m}$ constructed from $\CL_{\pm,m}$ is
labeled by an $n$-tuple of signs
$\sigma = (\sigma^1, \dotsc, \sigma^n) \in \{\pm\}^n$ and an $n$-tuple of
complex numbers $m = (m^1, \dotsc, m^n)$:
\begin{equation}
  \label{eq:M-trig}
  \vev{(\CM_{\sigma,m})_{i^1}^{i^{n+1}}}
  =
  \sum_{i^2, \dotsc, i^n}
  \prod_{s=1}^{n+1}
  e^{2\pi\iu b^s_{i^s}}
  \prod_{r=1}^n
  e^{\sigma^r \pi\iu(a^{r+1}_{i^{r+1}} - a^r_{i^r})}
  \ell_{m^r}(a^r, a^{r+1})_{i^r}^{i^{r+1}}
  \,.
\end{equation}
The corresponding transfer matrix $\CT_{\sigma,m}$ has the Wigner transform
\begin{equation}
  \label{eq:T-trig}
  \vev{\CT_{\sigma,m}}
  =
  \sum_{i^1, \dotsc, i^n}
  \prod_{r=1}^n e^{2\pi\iu b^r_{i^r}}
  e^{\sigma^r \pi\iu(a^{r+1}_{i^{r+1}} - a^r_{i^r})}
  \ell_{m^r}(a^r, a^{r+1})_{i^r}^{i^{r+1}}
  \,,
\end{equation}
with $a^{n+1} = a^1$, $i^{n+1} = i^1$.  Our claim is that these
quantities equal the vevs of Wilson--'t Hooft lines in $\CN = 2$
supersymmetric gauge theories.

\section{Wilson--'t Hooft lines as transfer matrices}
\label{sec:gauge}

In the previous section we defined the fundamental trigonometric
L-operators~\eqref{eq:fund-L} and calculated transfer matrices
constructed from them.  As explained in section~\ref{sec:intro}, these
transfer matrices are expected to have interpretations as Wilson--'t
Hooft lines in $\CN = 2$ supersymmetric gauge theories described by a
circular quiver.  In this section we verify this expectation by
computing the vevs of the corresponding Wilson--'t Hooft lines.

\subsection{Wilson--'t Hooft lines in
  $S^1 \times_\epsilon \R^2 \times \R$}

Consider a four-dimensional gauge theory whose gauge group is a
compact Lie group $G$ with Lie algebra $\gf$.  Choosing a maximal
torus $T \subset G$ with Lie algebra $\tf$, we let
$\rootl(\gf) \subset \tf^*$ and $\corootl(\gf) \subset \tf$ be the
root lattice and the coroot lattice of $\gf$, respectively.  Their
duals are the coweight lattice
$\coweightl(\gf) = \rootl(\gf)^\vee \subset \tf$ and the weight
lattice $\weightl(\gf) = \corootl(\gf)^\vee \subset \tf^*$.

An 't Hooft line is the worldline of a very heavy monopole, that is, a
nondynamical magnetically charged particle.  In the presence of an 't
Hooft line, the gauge field of the theory has a singularity at the
location of the monopole: in terms of the polar angle $\theta$ and the
azimuthal angle $\phi$ of the spherical coordinates centered at the
monopole, the gauge field behaves as
\begin{equation}
  \label{eq:monopole}
  A = \frac{\mcharge}{2} (1 - \cos\theta) \rmd\phi + \dotsb \,,
\end{equation}
where $\dotsb$ represents less singular terms.  (For simplicity we are
setting the gauge theory theta-angles to zero.)  The coefficient
$\mcharge$ is the magnetic charge of the monopole.  Different singular
gauge field configurations of the above form describe the same
monopole if their magnetic charges are related by gauge
transformation.  It follows that $\mcharge$ can be chosen from $\tf$,
and the choice is meaningful only up to the action of the Weyl group
$W(G)$ of $G$.

The above expression of $A$ is valid in a trivialization over a
coordinate patch that contains the point $\theta = 0$ of a two-sphere
surrounding the monopole.  At $\theta = \pi$, there is a ``Dirac
string'' which supports an unphysical magnetic flux.  For the Dirac
string to be invisible (or more precisely, for the gauge
transformation by $\exp(\iu\mcharge\phi)$ which allows us to go to the
coordinate patch containing $\theta = \pi$ to be well defined), we
must have
\begin{equation}
  \langle\mcharge, w\rangle \in \Z
\end{equation}
for every weight $w \in \tf^*$ of the representation of every field in
the theory.  This is simply the condition that the holonomy of $A$
around the point $\theta = \pi$ is trivial in the bundles of which the
fields are sections.  The theory always contains fields in the adjoint
representation, so $\mcharge$ belongs to the coweight lattice:%
\footnote{Further, $\mcharge$ belongs to the cocharacter lattice
  $\{v \in \tf \mid \exp(2\pi\iu v) = \id_G\}$, which is a sublattice
  of $\coweightl(\gf)$.  If we take $G$ to be the adjoint group, the
  cocharacter lattice coincides with $\coweightl(\gf)$.}
\begin{equation}
  \mcharge \in \coweightl(\gf)/W(G) \,. 
\end{equation}
Equivalently, $\mcharge$ is specified by an irreducible representation
of the Langlands dual ${}^L\gf$ of $\gf$.  In general, $\mcharge$ lies
in a sublattice of $\coweightl(\gf)/W(G)$ determined by the matter
content.

We can also consider heavy particles that carry both magnetic and
electric charges.  The worldline of such a dyon is called a Wilson--'t
Hooft line.  In the path integral formalism, a Wilson--'t Hooft line
is realized by an insertion of a Wilson line
\begin{equation}
  \label{eq:Wilson}
  \Tr_R P\exp\biggl(\iu\int_L A\biggr)
\end{equation}
and a singular boundary condition on the support $L$ of the line as
specified by the magnetic charge.  The prescribed
singularity~\eqref{eq:monopole} breaks the gauge symmetry to the
stabilizer $G_\mcharge$ of $\mcharge$, so $R$ is an irreducible
representation of $G_\mcharge$.  (More precisely, $R$ is an
irreducible representation of the stabilizer of $\mcharge$ in the
universal cover $\Gt$ of $G$~\cite{Kapustin:2005py}.)

The data specifying such a pair $(\mcharge,R)$ is actually the same as
a pair $(\mcharge, \echarge)$ of coweight $\mcharge$ and weight
$\echarge$ modulo the Weyl group action:
\begin{equation}
  (\mcharge, \echarge)
  \in
  \bigl(\coweightl(\gf) \times \weightl(\gf)\bigr)\big/W(G) \,. 
\end{equation}
As emphasized in~\cite{Kapustin:2005py}, this data has more information than
a pair of irreducible representations of $\gf$ and ${}^L\gf$.

In~\cite{Ito:2011ea}, the vevs of Wilson--'t Hooft lines in $\CN=2$
supersymmetric gauge theories on $S^1 \times_\eps \R^2 \times \R$ in
the Coulomb phase were computed via localization of the path integral.
The geometry $S^1 \times_\eps \R^2$ is a twisted product of $S^1$ and
$\R^2$, constructed from $[0,2\pi\beta] \times \R^2$ by the
identification $(2\pi\beta,z) \sim (0, e^{2\pi\iu\eps} z)$, where $z$ is
the complex coordinate of $\R^2 \iso \C$.  These Wilson--'t Hooft
lines wind around $S^1$, and are located at the origin of $\R^2$ and a
point in $\R$.  In order to preserve half of the eight supercharges,
they require the complex scalar field $\phi$ in the vector multiplet
to also have a singular behavior and replace the gauge field in the
Wilson line~\eqref{eq:Wilson} with $A + \iu\Re\phi$.  The vevs depend
holomorphically on parameters
\begin{equation}
  a \in \tf_\C \,,
  \qquad
  b \in \tf^*_\C \,,
\end{equation}
which are set by the values of the gauge field and the vector
multiplet scalar at spatial infinity.  Essentially, $a$ is given by
the holonomy around $S^1$ at infinity of the gauge field, while $b$ is
that of the dual gauge field.%
\footnote{Let $\theta_{\mathrm{e}}$ and $\theta_{\mathrm{m}}$ be the
  electric and magnetic theta-angles whose exponentials
  $e^{\iu\theta_{\mathrm{e}}}$ and $e^{\iu\theta_{\mathrm{m}}}$ are
  the electric and magnetic holonomies.  (The magnetic holonomy can be
  defined as the chemical potential for the magnetic charge in the
  path integral.)  Then, the parameters $a$, $b$ have semiclassical
  expansion
  \begin{equation}
    a = \frac{\theta_{\mathrm{e}}}{2\pi} + \iu\beta\Re\phi + \dotsb \,,
      \qquad
    b = \frac{\theta_{\mathrm{m}}}{2\pi} - \frac{4\pi\iu\beta}{g^2} \Im\phi
        + \iu\frac{\vartheta}{2\pi} \beta\Re\phi + \dotsb \,,
  \end{equation}
  where $g$ is the gauge coupling and $\vartheta$ is the gauge theory
  theta-angle.  These parameters are complexified Fenchel--Nielsen
  coordinates on the Seiberg--Witten moduli space and receive
  nonperturbative corrections (indicated above by the ellipses) which
  are known~\cite{Gaiotto:2008cd}.  See~\cite{Brennan:2019hzm} for a
  recent discussion on the nonperturbative corrections in the present
  context.}

The vev of a Wilson line $W_R$ in representation $R$ is simply given
by the classical value of the holonomy:
\begin{equation}
  \label{eq:vev-W}
    \vev{W_R} = \Tr_R e^{2\pi \iu a} \,.
\end{equation}

The vevs of 't Hooft lines are much more involved.  For an 't Hooft
line $T_\mcharge$ with magnetic charge $\mcharge$, the vev takes the form
\begin{equation}
  \label{eq:vev-T}
  \vev{T_\mcharge}
  =
  \sum_{\substack{v \in \corootl(\gf) + \mcharge \\ \|v\| \leq \|\mcharge\|}}
  e^{2\pi \iu \langle v, b\rangle}
  Z_{\text{1-loop}}(a,m,\eps; v) Z_{\text{mono}}(a,m,\eps; \mcharge,v) \,,
\end{equation}
where $m$ collectively denotes complex mass parameters.  The summation
over the coweights $v$ in the shifted coroot lattice
$\corootl + \mcharge$ accounts for the so-called ``monopole
bubbling,'' a phenomenon in which smooth monopoles are absorbed by the
't Hooft line and screen the magnetic charge.  The norm $\|v\|$ with
respect to a Killing form is bounded by $\|\mcharge\|$, so this is a
finite sum.  The first two factors in the summand are the classical
action and the one-loop determinant in the screened monopole
background, respectively.  The last factor is the nonperturbative
contributions coming from degrees of freedom trapped on the 't Hooft
line due to monopole bubbling.

Suppose that the theory under consideration consists of a vector
multiplet and $N_F$ hypermultiplets in representations $R_f$ with mass
parameters $m_f$, $f = 1$, $\dotsc$, $N_F$.  The one-loop determinant
$Z_{\text{1-loop}}$ is then the product of the contributions from the
vector multiplet and the hypermultiplets:
\begin{equation}
  Z_{\text{1-loop}}(a,m,\eps; v)
  =
  Z_{\text{1-loop}}^{\mathrm{vm}}(a,\eps; v)
  \prod_{f=1}^{N_F} Z_{\text{1-loop}}^{\mathrm{hm}, R_f}(a,m_f,\eps; v) \,.
\end{equation}
The two functions are given by
\begin{align}
  Z_{\text{1-loop}}^{\mathrm{vm}}(a,\eps; v)
  &=\prod_{\alpha \in \Phi(\gf)} \prod_{k=0}^{|\langle v, \alpha\rangle| - 1}
    \sin^{-\frac12}
    \biggl(\pi\langle a, \alpha\rangle + \pi\Big(\frac12 |\langle v, \alpha\rangle| - k\Bigr)\eps\biggr) \,,
  \\  
  Z_{\text{1-loop}}^{\mathrm{hm}, R}(a,m,\eps; v)
  &=\prod_{w \in P(R)}\prod_{k=0}^{|\langle v, w\rangle| - 1}
    \sin^{\frac12}\biggl(\pi \langle a, w\rangle - \pi m
    + \pi\Bigl(\frac12|\langle v, w\rangle| - \frac12 - k\Bigr)\eps\biggr) \,.
\end{align}
Here, $\Phi(\gf)$ is the set of roots of $\gf$ and $P(R)$ is the set
of weights of $R$.

The factor $Z_{\text{mono}}$ is subtle.  The original computation
in~\cite{Ito:2011ea} did not give an answer that completely matches
predictions from the AGT correspondence.  The subtleties have been
addressed in subsequent works~\cite{Brennan:2018yuj, Brennan:2018moe,
  Brennan:2018rcn, Assel:2019iae} but not resolved in full generality.

Fortunately, for Wilson--'t Hooft lines that are of interest to us,
the screened magnetic charges are in the same $W(G)$-orbit as
$\mcharge$.  The corresponding contributions are therefore obtained by
the $W(G)$-action from the perturbative term, for which $v = \mcharge$
and $Z_{\text{mono}} = 1$.

To our knowledge, a formula for the vevs of dyonic Wilson--'t Hooft
lines generalizing the expressions~\eqref{eq:vev-W}
and~\eqref{eq:vev-T} has not been derived.  Nevertheless, for the same
reason as mentioned, we can calculate the vev of a relevant Wilson--'t
Hooft line by first writing down its perturbative contribution, which
is simply the product of the perturbative vevs of the corresponding
purely electric and purely magnetic lines, and then summing over the
contributions from the nonperturbative sectors related by the
$W(G)$-action.

\subsection{Transfer matrices from circular quiver theories}
\label{sec:circular-quiver-theory}

The Wilson--'t Hooft line that corresponds to the transfer
matrix~\eqref{eq:T-trig} is one in an $\CN = 2$ supersymmetric gauge
theory that is described by a circular quiver with $n$ nodes:
\begin{equation}
  \begin{tikzpicture}[scale=0.8, baseline=(x.base)]
    \node (x) at (0,0) {\vphantom{x}};
    
    \draw[dotted] (163:1) arc (163:197:1);
    \draw (-155:1) arc (-155:155:1);

    \node[gnode] at (0:1) {$N$};
    \node[gnode] at (60:1) {$N$};
    \node[gnode] at (120:1) {$N$};
    \node[gnode] at (240:1) {$N$};
    \node[gnode] at (300:1) {$N$};
  \end{tikzpicture}
  \quad .
\end{equation}
Each node represents a vector multiplet for an $\SU(N)$ gauge group,%
\footnote{More precisely, the gauge group is a product of $\PSU(N)$.}
and each edge a hypermultiplet that transforms in the bifundamental
representation under the gauge groups of the nodes it connects.

Let us first consider the case in which the quiver consists of a
single node and a single edge.  In this case, the gauge group
$G = \SU(N)$ and the only hypermultiplet is in the adjoint
representation.  This theory is known as $\CN = 2^*$ theory.

The roots of $\gf = \suf_N$ are
$\alpha_{ij} = E_{ii}^* - E_{jj}^* = h_i - h_j$, $i \neq j$.  The
positive roots are $\alpha_{ij}$, $i < j$, and the simple roots are
$\alpha_i = \alpha_{i,i+1}$, $i = 1, \,, \dotsc,\, N-1$.  The
fundamental coweights are
$\omega^\vee_i = (\alpha_i^\vee)^* = \sum_{j=1}^i h^\vee_j$, with
\begin{equation}
  h^\vee_i = E_{ii} - \frac{1}{N} \sum_{j=1}^N E_{jj} \,.
\end{equation}
The various lattices are
\begin{equation}
  \rootl = \bigoplus_{i=1}^{N-1} \Z \alpha_i \,,
  \qquad
  \corootl = \bigoplus_{i=1}^{N-1} \Z \alpha^\vee_i \,,
  \qquad
  \weightl = \bigoplus_{i=1}^{N-1} \Z \omega_i \,,
  \qquad
  \coweightl = \bigoplus_{i=1}^{N-1} \Z \omega^\vee_i \,.
\end{equation}
We recall that $\alpha^\vee_i$ are the simple coroots and
$\omega_i = (\alpha^\vee_i)^*$ are the fundamental weights.

For $\CN = 2^*$ theory with $G = \SU(N)$, minimal magnetic charges are
$\mcharge = \omega^\vee_1 = h^\vee_1$ and
$\mcharge = \omega^\vee_{N-1} = -h^\vee_N$.  These magnetic charges are the
highest weights of the fundamental representation and the
antifundamental representation of ${}^L\suf_N \iso \suf_N$,
respectively.

Let us consider the 't Hooft line with $\mcharge = h^\vee_1$.  The vev
of this 't Hooft line is expressed as a sum over the screened magnetic
charges $v = h^\vee_1$, $h^\vee_2$, $\dotsc$, $h^\vee_N$.  The term
for $v = h^\vee_1$ is the perturbative contribution and given by
\begin{equation}
  e^{2\pi\iu b_1}
  \prod_{j=2}^N
  \sin^{-\frac12}\Bigl(\pi a_{1j} + \frac12 \pi\eps\Bigr)
  \sin^{-\frac12}\Bigl(\pi a_{j1} + \frac12 \pi\eps\Bigr)
  \sin^{\frac12}(\pi a_{1j} - \pi m)
  \sin^{\frac12}(\pi a_{j1} - \pi m) \,,
\end{equation}
where $a_i = \langle a, h_i\rangle$,
$b_i = \langle h^\vee_i, b\rangle$, $a_{ij} = a_i - a_j$ and $m$ is
the mass of the adjoint hypermultiplet.  The other terms are related
to this perturbative term by the Weyl group action which permutes
$(h^\vee_1, \dotsc, h^\vee_N)$, so we find
\begin{equation}
  \vev{T_{h^\vee_1}}
  =
  \sum_{i=1}^N
  e^{2\pi\iu b_i}
  \prod_{j (\neq i)}
  \biggl(
  \frac{\sin\pi(a_{ij} - m) \sin\pi(a_{ji} - m)}
       {\sin\pi(a_{ij} - \frac12\eps) \sin\pi(a_{ji} - \frac12\eps)}
  \biggr)^{\frac12} \,.
\end{equation}
The vev of $T_{-h^\vee_N}$ is obtained from $\vev{T_{h^\vee_1}}$ by
the replacement $b_i \to -b_i$.

Now, let us turn to a circular quiver with $n$ nodes.  For this
theory, we have $G = \SU(N)^n$ and
$\coweightl(\gf) = \coweightl(\suf_N)^{\oplus n}$.  We consider the 't
Hooft line with
\begin{equation}
  \label{eq:B-circular}
  \mcharge =  h^\vee_1 \oplus \dotsb \oplus h^\vee_1 \,,
\end{equation}
charged equally under the $\SU(N)$ factors of $G$.  This time, the
summation is over all coweights of the form
$v = h^\vee_{i^1} \oplus \dotsb \oplus h^\vee_{i^n}$.  The perturbative term,
for which $i^1 = \dotsb = i^n = 1$, is given by
\begin{multline}
  \label{eq:T-circ-pert}
  \prod_{r=1}^n
  e^{2\pi\iu b^r_1}
  \prod_{j=2}^N
  \sin^{-\frac12}\Bigl(\pi a^r_{1j} + \frac12 \pi\eps\Bigr)
  \sin^{-\frac12}\Bigl(\pi a^r_{j1} + \frac12 \pi\eps\Bigr)
  \\
  \times
  \sin^{\frac12}\bigl(\pi(a^r_j - a^{r+1}_1) - \pi m^r \bigr)
  \sin^{\frac12}\bigl(\pi(a^r_1 - a^{r+1}_j) - \pi m^r\bigr)
  \,.
\end{multline}
The superscript $r$ refers to the $r$th $\SU(N)$ factor of $G$, with
$a^{n+1} = a^1$.  Collecting the contributions from the other
coweights, we get
\begin{equation}
  \vev{T_{h^\vee_1 \oplus \dotsb \oplus h^\vee_1}}
  =
  \sum_{i^1, \dotsc, i^n}
  \prod_{r=1}^n
  e^{2\pi\iu b^r_{i^r}}
  \ell_{m^r}(a^r, a^{r+1})_{i^r}^{i^{r+1}} \,,
\end{equation}
where we have used the functions~\eqref{eq:ell}.

Comparing this expression with the Wigner transform~\eqref{eq:T-trig}
of the trigonometric transfer matrix $\CT_{\sigma,m}$, we see
\begin{equation}
  \vev{T_{h^\vee_1 \oplus \dotsb \oplus h^\vee_1}}
  =
  \vev{\CT_{(+,\dotsc,+), m}}
  =
  \vev{\CT_{(-,\dotsc,-), m}}
\end{equation}
under the obvious identification of parameters.

In order to reproduce $\vev{\CT_{\sigma,m}}$ for a general choice of
the signs $\sigma$, we add to the 't Hooft line the electric charge
\begin{equation}
  \label{eq:E-circular}
  \echarge
  = \sum_{r=1}^n \sigma^r \frac12(h^{r+1}_1 - h^r_1)
  = \sum_{r=1}^n (\sigma^r 1 - \sigma^{r+1} 1)  \frac12 h^{r+1}_1 \,.
\end{equation}
This electric charge is in a sense a minimal one that is compatible
with the Dirac--Schwinger--Zwanziger quantization condition for
locality: the charges $(\mcharge,\echarge)$ and
$(\mcharge',\echarge')$ of two dyons must satisfy
$\langle \mcharge, \echarge'\rangle - \langle \mcharge',
\echarge\rangle \in \Z$.  In section~\ref{sec:Toda}, we will see the
geometric meaning of this ``minimality'' in connection with the AGT
correspondence.

The magnetic charge~\eqref{eq:B-circular} breaks the gauge group to
$\mathrm{S}(\U(1) \times \U(N-1))^n$, and we are turning on a Wilson
line that is charged under the $\U(1)$ factors with charges
proportional to $(\sigma^r 1 - \sigma^{r+1} 1)/2$.  The Wilson line
multiplies the perturbative term~\eqref{eq:T-circ-pert} by the phase
factor
\begin{equation}
  \prod_{r=1}^n e^{\sigma^r \pi\iu \langle a, h^{r+1}_1 - h^r_1\rangle}
  =
  \prod_{r=1}^n e^{\sigma^r \pi\iu (a^{r+1}_1 - a^r_1)} \,.
\end{equation}
Hence, the term with $v = h^\vee_{i^1} \oplus \dotsb \oplus h^\vee_{i^n}$ gets
the phase factor
$e^{\sigma^r \pi\iu (a^{r+1}_{i^{r+1}} - a^r_{i^r})}$, and the vev of
this Wilson--'t Hooft line matches the Wigner transform of
$\CT_{\sigma,m}$.

\subsection{Monodromy matrices from linear quiver theories}

We have considered the Wilson--'t Hooft lines in the circular quiver
theory and showed that their vevs match the Wigner transforms of the
trigonometric transfer matrices.  What correspond to the monodromy
matrices then?  In view of the fact that summing over the weights of
the representation $V = \C^N$ in the integrable model amounts to
summing over the different screened magnetic charges, natural
candidates are Wilson--'t Hooft lines in a theory described by a
linear quiver with $n + 1$ nodes:
\begin{equation}
  \label{eq:linear-q}
  \begin{tikzpicture}[baseline=(x.base)]
    \node (x) at (0,0) {\vphantom{x}};

    \node[fnode] (1) at (0,0) {$N$};
    \node[gnode] (2) at (1,0) {$N$};
    \node[gnode] (3) at (2,0) {$N$};
    \node[gnode] (n) at (3.6,0) {$N$};
    \node[fnode] (n+1) at (4.6,0) {$N$};

    \draw (1) -- (2) -- (3) -- +(0:0.5);
    \draw[dotted] (2.5,0) -- (3.1,0);
    \draw (n) -- +(180:0.5);
    \draw (n) -- (n+1);
  \end{tikzpicture}
  \quad .
\end{equation}
The leftmost and the rightmost nodes represent $\SU(N)$ flavor groups,
which are not gauged.

In particular, we expect that the fundamental trigonometric
L-operators~\eqref{eq:fund-L} arise from the vevs of Wilson--'t Hooft
lines of the theory of a bifundamental hypermultiplet:
\begin{equation}
  \label{eq:two-node-q}
  \begin{tikzpicture}[baseline=(x.base)]
    \node (x) at (0,0) {\vphantom{x}};
    
    \node[fnode] (1) at (0,0) {$N$};
    \node[fnode] (2) at (1,0) {$N$};
    \draw (1) -- (2);
  \end{tikzpicture}
  \quad .
\end{equation}
Let us see if this is the case.

We introduce nondynamical vector multiplets for the $\SU(N)$ flavor
groups, and consider the Wilson--'t Hooft lines with magnetic charge
\begin{equation}
  \mcharge = h^\vee_i \oplus h^\vee_j
\end{equation}
and electric charges
\begin{equation}
  \echarge =  \mp\frac12 h_i \oplus \pm\frac12 h_j\,.
\end{equation}
Note that the electric charges are fractional.  The vevs of these
Wilson--'t Hooft lines are
\begin{equation}
  e^{2\pi\iu (b^1_i + b^2_j)}
  e^{\pm\pi\iu (a^2_j - a^1_i)}
  \prod_{k (\neq i)}
  \prod_{l (\neq j)}
  \bigl(\sin\pi(a^1_k - a^2_j - m)
  \sin\pi(a^1_i - a^2_l - m)\bigr)^{\frac12} \,.
\end{equation}
The vevs do not quite match the Wigner transforms~\eqref{eq:fund-L-ij}
of $(\CL_{\pm,m})_i^j$.  They differ by the factor in the denominator
of the function~\eqref{eq:ell}.

This factor is the one-loop determinant associated with the first
node; it would have been present had the $\SU(N)$ flavor group been
gauged and the vector multiplet been dynamical.  From the gauge theory
point of view, it is natural to think of this factor as a weight
accompanying the summation over the screened magnetic charges.  On the
integrable system side, we could as well omit the denominator in
question from the definitions of the L-operators and adopt the
convention that the same weight is included when operators are
multiplied within $V$.  The L-operators would still satisfy the RLL
relation.

To get the monodromy matrix~\eqref{eq:M-trig}, we take $n$ L-operators
and multiply them inside $V$.  The gauge theory counterpart of this
operation is to connect $n$ copies of the two-node
quiver~\eqref{eq:two-node-q}, in the presence of appropriate
Wilson--'t Hooft lines of the type considered above, by identifying
and gauging flavor nodes.  This produces the $n+1$ node linear
quiver~\eqref{eq:linear-q} and the Wilson--'t Hooft lines with
magnetic charge
\begin{equation}
  \label{eq:B-linear}
  \mcharge
  =
  h^\vee_{i^1} \oplus h^\vee_1 \oplus h^\vee_1 \oplus \dotsb \oplus h^\vee_1
  \oplus h^\vee_{i^{n+1}}
\end{equation}
and electric charge
\begin{equation}
  \echarge
  =
  \sigma^1 \frac12 (h_1^2 - h_{i^1}^1)
  + \sum_{r=2}^{n-1} \sigma^r \frac12 (h^{r+1}_1 - h^r_1)
  + \sigma^n \frac12 (h_{i^{n+1}}^{n+1} - h_1^n) \,.
\end{equation}
The vev of this Wilson--'t Hooft line reproduces the Wigner
transform~\eqref{eq:M-trig}, except that a factor corresponding to the
one-loop determinant for the vector multiplet for the first node is
missing.

\subsection{Other representations}

The magnetic charge~\eqref{eq:B-circular} of the above Wilson--'t
Hooft lines is the highest weight of the representation
$(\C^N)^{\oplus n}$ of the Langlands dual
${}^L\gf_\C \iso \slf_N^{\oplus n}$ of $\gf_\C$.  The corresponding
transfer matrix~\eqref{eq:T-trig} is represented graphically as $n$
double lines intersected by a single solid loop carrying the
representation $V = \C^N$, as shown in figure~\ref{fig:spin-chain-T}.
The $n$ regions sandwiched between double lines correspond to the $n$
copies of $\slf_N$.

Both sides of the correspondence have a generalization in which the
vector representation $\C^N$ is replaced by another representation $R$
of $\slf_N$.  On the gauge theory side, we can change the magnetic
charge of the Wilson--'t Hooft lines to the highest weight $\lambda_R$
of $R^{\oplus n}$ while keeping the electric charges intact.  On the
integrable system side, the counterpart of this operation is the
fusion procedure, which allows one to construct a solid line in an
arbitrary finite-dimensional representation of $\slf_N$ from a
collection of solid lines in the vector representation, with the
spectral parameters suitably adjusted.

We naturally expect that the vev of the Wilson--'t Hooft line with
magnetic charge $\mcharge = \lambda_R^{\oplus n}$ is equal to the
Wigner transform of a transfer matrix constructed from L-operators in
representation $R$, obtained by fusion from the
L-operators~\eqref{eq:fund-L} in the vector representation.

For $n = 1$ and $R = \wedge^k \C^N$, this equality can be verified
from known results.  In this case, the transfer matrix is the
trigonometric limit of Ruijsenaars' difference
operator~\cite{Ruijsenaars:1986pp}
\begin{equation}
  \sum_{\substack{I \subset \{1, \dotsc, N\} \\ |I| = k}}
  \Delta_I^{\frac12}
  \prod_{\substack{i \in I \\ j \notin I}}
  \sqrt{
    \frac{\theta_1(a_{ji} - m) \theta_1(a_{ij} - m)}
         {\theta_1(a_{ji} - \frac12 \eps) \theta_1(a_{ij} - \frac12 \eps)}
  }
  \Delta_I^{\frac12} \,,
  \qquad
  \Delta_I = \prod_{i \in I} \Delta_i \,,
\end{equation}
and is related to the Macdonald operator by a similarity
transformation~\cite{MR1463830}.  On the other hand, the exterior
power $\wedge^k \C^N$ being a minuscule representation (that is, all
weights are related by the action of the Weyl group), the vev of the
't Hooft line with
$\mcharge = \omega^\vee_k = h^\vee_1 + \dotsb + h^\vee_k$ in
$\CN = 2^*$ theory can be computed from the perturbative term:
\begin{equation}
  \vev{T_{\omega^\vee_k}}
  =
  \sum_{\substack{I \subset \{1, \dotsc, N\} \\ |I| = k}}
  \prod_{\substack{i \in I \\ j \notin I}}
  e^{2\pi\iu b_i}
  \biggl(
  \frac{\sin\pi(a_{ij} - m) \sin\pi(a_{ji} - m)}
       {\sin\pi(a_{ij} - \frac12\eps) \sin\pi(a_{ji} - \frac12\eps)}
  \biggr)^{\frac12} \,.
\end{equation}
(For $\CN = 2^*$ theory the choice of the signs $\sigma = \pm$ is
irrelevant.)  The vev matches the Wigner transform of the
trigonometric Ruijsenaars operator.

\section{Transfer matrices from Verlinde operators}
\label{sec:Toda}

We have computed the vevs of a class of Wilson--'t Hooft lines in
$\CN = 2$ supersymmetric gauge theories described by a circular
quiver, and found that they match the Wigner transforms of transfer
matrices constructed from the fundamental trigonometric L-operators.
In this section, we show that these transfer matrices can also be
identified with Verlinde operators in Toda theory on a punctured
torus.  The result is in keeping with the relation proposed
in~\cite{Ito:2011ea} based on the AGT
correspondence~\cite{Alday:2009aq} between Toda theory and $\CN = 2$
supersymmetric field theories.

\subsection{Verlinde operators and Wilson--'t Hooft lines}

The AGT correspondence originates from six-dimensional $\CN = (2,0)$
supersymmetric field theory, of type $A_{N-1}$ in our case, placed on
$S^4_\bs \times C_{g,n}$.  Here, $S^4_\bs$ is an ellipsoid, defined as
a submanifold of $\R^5$ by the equation
\begin{equation}
  (x^1)^2
  + \bs^{-2} \bigl((x^2)^2 + (x^3)^2\bigr)
  + \bs^2 \bigl((x^4)^2 + (x^5)^2\bigr)
  = r^2 \,,
\end{equation}
and $C_{g,n}$ is a Riemann surface of genus $g$ with $n$ punctures.
With partial topological twisting along $C_{g,n}$, this system
preserves eight of the sixteen supercharges of $\CN = (2,0)$
supersymmetry in six dimensions.

In the limit in which $C_{g,n}$ shrinks to a point, the
six-dimensional theory reduces to a four-dimensional $\CN = 2$
supersymmetric field theory on $S^4_\bs$, whose gauge and matter
contents are determined by the choice of a pants decomposition of
$C_{g,n}$ and boundary conditions at the
punctures~\cite{Gaiotto:2009we, Gaiotto:2009hg}.  The theories
discussed in section~\ref{sec:gauge} can all be obtained in this way.
If one instead integrates out the modes along $S^4_\bs$, one is left
with $A_{N-1}$ Toda theory on $C_{g,n}$ with central charge
\begin{equation}
  c = 1 + 6q^2 \,,
  \qquad
  q = \bs + \bs^{-1} \,,
\end{equation}
with vertex operators $V_{\beta^r}$, $r = 1$, $\dotsc$, $n$, inserted
at the punctures.  According to the AGT correspondence, the partition
function of the theory on $S^4_\bs$ equals the correlation function of
Toda theory on $C_{g,n}$:
\begin{equation}
  \vev{1}_{S^4_\bs} = \biggvev{\prod_r V_{\beta^r}}_{C_{g,n}} \,.
\end{equation}

Let the vertex operators at the punctures be primary fields
$V_{\beta^r}$, $r = 1$, $\dotsc$, $n$, labeled by momenta
$\beta^r \in \hf^*$ valued in the dual of the Cartan subalgebra $\hf$
of $\slf_N$.  Given a pants decomposition of $C_{g,n}$, the Toda
correlation function takes the form
\begin{equation}
  \biggvev{\prod_r V_{\beta^r}}_{C_{g,n}}
  =
  \int [\rmd\alpha] \,
  \CC(\alpha; \beta)
  \overline{\CF(\alpha; \beta)} \CF(\alpha; \beta) \,,
\end{equation}
where $[\rmd\alpha]$ is a measure of integration over the set
$\alpha = \{\alpha^1, \dotsc, \alpha^{3g - 3 + n}\}$ of momenta
assigned to the internal edges of the pants decomposition,
$\beta = \{\beta^1, \dotsc, \beta^n\}$ is the set of momenta assigned
to the external edges, $\CC(\alpha; \beta)$ is the product of relevant
three-point functions, and $\CF(\alpha; \beta)$ is the corresponding
conformal block which is a meromorphic function of $\alpha$ and
$\beta$.

On the gauge theory side, $\CC(\alpha; \beta)$ is interpreted as the
product of the classical and the one-loop contributions to the
partition function on $S^4_\bs$, whereas $\CF(\alpha; \beta)$ and
$\overline{\CF(\alpha; \beta)}$ represent the nonperturbative
contributions from instantons localized at the two poles at
$x^2 = x^3 = x^4 = x^5 = 0$.  The internal momenta $\alpha$ are
related to the zero modes $\as$ of scalar fields in the vector
multiplets by
\begin{equation}
  \alpha = Q + \iu\as \,,
\end{equation}
and the external momenta $\beta$ are identified with mass parameters
for matter multiplets.

To incorporate Wilson--'t Hooft lines in the gauge theory, one
introduces Verlinde loop operators in the Toda theory.  We will
explain the construction of relevant Verlinde operators in concrete
examples.  For the moment, it suffices to say that they are specified
by a momentum of the form $\mu = -\bs\lambda$ and a one-cycle $\gamma$
in $C_{g,n}$, where $\lambda$ is the highest weight of a
representation of $\slf_N$.%
\footnote{More generally, the momentum takes the form
  $\mu = -\bs\lambda_1 - \bs^{-1} \lambda_2$, where $\lambda_1$,
  $\lambda_2$ are the highest weights of a pair of representations of
  $\slf_N$.  The corresponding Wilson--'t Hooft line is a
  superposition of lines wrapping $S^1_\bs$ and another circle
  $S^1_{\bs^{-1}}$ where $x^2 = x^3 = 0$.}
In the presence of a Verlinde operator $\Phi_\mu(\gamma)$, the Toda
correlation function is modified to
\begin{equation}
  \biggvev{\Phi_\mu(\gamma) \prod_r V_{\beta^r}}_{C_{g,n}}
  =
  \int [\rmd\alpha] \,
  \CC(\alpha; \beta) \overline{\CF(\alpha; \beta)}
  \bigl(\Phi_\mu(\gamma) \cdot \CF(\alpha; \beta)\bigr) \,.
\end{equation}
The AGT correspondence asserts \cite{Alday:2009fs, Drukker:2009id}
that this is equal to the vev of a Wilson--'t Hooft line
$T_{\mu,\gamma}$ winding around a circle $S^1_\bs$ where
$x^4 = x^5 = 0$ (at $x^1 = 0$, say):
\begin{equation}
  \vev{T_{\mu,\gamma}}_{S^4_\bs}
  =
  \biggvev{\Phi_\mu(\gamma) \prod_r V_{\beta^r}}_{C_{g,n}} \,.
\end{equation}

It turns out that $\Phi_\mu(\gamma)$ acts on conformal blocks as a
difference operator shifting the internal momenta $\alpha$, just as
Wilson--'t Hooft lines in $\CN = 2$ supersymmetric gauge theories on
$S^1 \times_\eps \R^2 \times \R$ shift Coulomb branch parameters.  Indeed, it
was argued in~\cite{Ito:2011ea} that if one defines the modified Verlinde
operator
\begin{equation}
  \label{eq:c-verlinde}
  \CL_\mu(\gamma)
  =
  \CC(\alpha; \beta)^{\frac12}
  \Phi_\mu(\gamma) \CC(\alpha; \beta)^{-\frac12} \,,
\end{equation}
then its Wigner transform is equal to the vev of the Wilson--'t Hooft
line in the theory on $S^1 \times_\eps \R^2 \times \R$, up to an appropriate
identification of parameters:
\begin{equation}
  \vev{T_{\mu,\gamma}}_{S^1 \times_\eps \R^2 \times \R}
  =
  \vev{\CL_\mu(\gamma)} \,.
\end{equation}
Therefore, we expect that for suitable choices of $C_{g,n}$, $\beta$,
$\mu$ and $\gamma$, the modified Verlinde operator $\CL_\mu(\gamma)$
coincides with a transfer matrix constructed from the trigonometric
L-operator.

\subsection{Verlinde operators on a punctured torus}

To reproduce the transfer matrix~\eqref{eq:T-trig}, we consider Toda
theory on an $n$-punctured torus $C_{1,n}$ and insert vertex operators
$V_{\beta^r}$ with
\begin{equation}
  \beta^r
  = -N\Bigl(\frac{q}{2} + \iu\ms^r\Bigr) h_N \,.
\end{equation}
The corresponding four-dimensional theory on $S^4_\bs$ is the one
described by an $n$-node circular quiver, which we studied in
section~\ref{sec:circular-quiver-theory}.  The parameter $\ms^r$ is
the mass of the bifundamental hypermultiplet between the $r$th and
$(r+1)$th nodes.

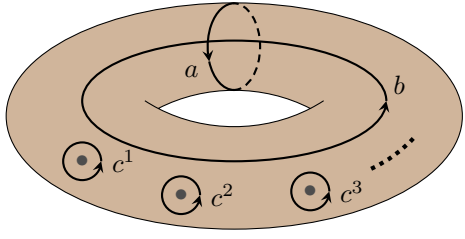
\begin{figure}
  \centering
  \begin{tikzpicture}[font=\small]

    \draw[fill={rgb,255:red,208; green,181; blue,155},even odd rule]
    (0,0) ellipse (3 and 1.5) 
    (-1,0.1) arc (120:60:2 and 1.8) arc (-60:-120:2 and 1.8);

    \draw (-1,0.1) arc(-120:-126:2 and 1.8) (1,0.1) arc(-60:-54:2 and 1.8);

    \draw[thick, ->] (2,0.2) arc (0:360:2 and 0.8) node[above right=-2pt] {$b$};

    \draw[thick, ->] (0,1.495)
    node[shift={(-16pt,-0.9)}] {$a$} arc (90:200:0.35 and 0.575);
    \draw[thick] (-0.33,0.7235) arc (200:270:0.35 and 0.575);
    \draw[thick, densely dashed] (-0.01,0.345) arc (270:450:0.35 and 0.575);


    \node[black!70] (p3) at (1,-1) {$\bullet$};
    \node[black!70] (p2) at (-0.7,-1.05) {$\bullet$};
    \node[black!70] (p1) at (-2,-0.6) {$\bullet$};

    \draw[ultra thick, dotted] (1.8,-0.7) arc (-65:-42:1.8);

    \draw[thick, ->] ($(p1)+(0.25,0)$) arc (0:360:0.25) node[right] {$c^1$};
    \draw[thick, ->] ($(p2)+(0.25,0)$) arc (0:360:0.25) node[right] {$c^2$};
    \draw[thick, ->] ($(p3)+(0.25,0)$) arc (0:360:0.25) node[right] {$c^3$};
  \end{tikzpicture}
  \caption{One-cycles on a punctured torus. The cycle $c^r$ goes
    around the $r$th puncture.}
  \label{fig:cycles-on-torus}
\end{figure}

To this setup we introduce the Verlinde operator $\Phi_\mu(\gamma)$
with
\begin{equation}
  \mu = -\bs\omega_1 = -\bs h_1
\end{equation}
and $\gamma$ being a cycle $\gamma_\sigma$ specified by an $n$-tuple
of signs $\sigma \in \{\pm\}^n$.  If $b$ and $c^r$ are the cycles
shown in figure~\ref{fig:cycles-on-torus}, then
\begin{equation}
  \gamma_\sigma
  =
  b + \sum_r \frac{1 - \sigma^r1}{2} c^r \,.
\end{equation}
In other words, the curve $\gamma_\sigma$ passes ``above'' or
``below'' the $r$th puncture depending on whether $\sigma^r = +$ or
$-$.  In the gauge theory, this operator corresponds to the Wilson--'t
Hooft line with magnetic charge~\eqref{eq:B-circular} and electric
charge~\eqref{eq:E-circular}.

Let us explain the construction of this Verlinde operator step by
step, following the treatment in~\cite{Gomis:2010kv}.  To this end,
it is convenient to represent the conformal block graphically as
\begin{equation}
\begin{tikzpicture}[scale=0.5, baseline=(x.base)]
  \node (x) at (0,0) {\vphantom{x}};
  
  \draw [ultra thick, ->-=0.5] (0,0) circle [radius=1.5];

  \draw [ultra thick, -<-=0.6] (60:1.5) -- +(60:1)
  node[shift={(60:8pt)}] {$\beta^1$};

  \draw [ultra thick, -<-=0.6] (0:1.5) -- +(0:1)
  node[shift={(0:8pt)}] {$\beta^2$};

  \draw [ultra thick, -<-=0.6] (-60:1.5) -- +(-60:1)
  node[shift={(-60:6pt)}] {$\beta^3$};

  \draw [ultra thick, -<-=0.6] (-120:1.5) -- +(-120:1)
  node[shift={(-120:8pt)}] {$\beta^4$};

  \draw [ultra thick, dotted] (-160:2) arc (-160:-200:2);

  \draw [ultra thick, -<-=0.6] (-240:1.5) -- +(-240:1)
  node[shift={(-240:6pt)}] {$\beta^n$};

  \node[shift={(90:-7pt)}] at (90:1.5) {$\alpha^1$};
  \node[shift={(30:-8pt)}] at (30:1.5) {$\alpha^2$};
  \node[shift={(-30:-7pt)}] at (-30:1.5) {$\alpha^3$};
  \node[shift={(-90:-8pt)}] at (-90:1.5) {$\alpha^4$};
\end{tikzpicture}
\,.
\end{equation}
The internal momenta are $\alpha^r$, $r = 1$, $\dotsc$, $n+1$, with
$\alpha^{n+1} = \alpha^1$.

The first step is to insert the identity operator between $\beta^n$ and
$\beta^1$, and resolve it into the chiral vertex operators $V_{-\bs h_1}$
and $V_{\bs h_N}$ by fusion.  This step gives the equality
\begin{equation}
\begin{tikzpicture}[scale=0.5, baseline=(x.base)]
  \node (x) at (0,0) {\vphantom{x}};

  \draw [ultra thick, ->-=0.85] (0,0) circle [radius=1.5];

  \draw [ultra thick, -<-=0.6] (60:1.5) -- +(60:1)
  node[shift={(60:8pt)}] {$\beta^1$};

  \draw [ultra thick, -<-=0.6] (0:1.5) -- +(0:1)
  node[shift={(0:8pt)}] {$\beta^2$};

  \draw [ultra thick, dotted] (-40:2) arc (-40:-80:2);

  \draw [ultra thick, -<-=0.6] (-120:1.5) -- +(-120:1)
  node[shift={(-120:8pt)}] {$\beta^n$};

  \draw [thick, densely dashed, -<-=0.6] (150:1.5) --
  node[shift={(60:6pt)}] {$0$} +(150:1);

  \draw [thick, -<-=0.6] (150:2.5) -- +(90:1)
  node[shift={(90:6pt)}] {$-\bs h_1$};

  \draw [thick, -<-=0.6] (150:2.5) -- +(210:1)
  node[shift={(210:8pt)}] {$\bs h_N$};

  \node[shift={(105:-7pt)}] at (105:1.5) {$\alpha^1$};
  \node[shift={(30:-8pt)}] at (30:1.5) {$\alpha^2$};
  \node[shift={(-165:-14pt)}] at (-165:1.5) {$\alpha^{n+1}$};
\end{tikzpicture}
=
\sum_{i^1} F_{i^1}
\begin{tikzpicture}[scale=0.5, baseline=(x.base)]
  \node (x) at (0,0) {\vphantom{x}};

  \draw [ultra thick, ->-=0.85] (0,0) circle [radius=1.5];

  \draw [ultra thick, -<-=0.6] (60:1.5) -- +(60:1)
  node[shift={(60:8pt)}] {$\beta^1$};

  \draw [ultra thick, -<-=0.6] (0:1.5) -- +(0:1)
  node[shift={(0:8pt)}] {$\beta^2$};

  \draw [ultra thick, dotted] (-40:2) arc (-40:-80:2);

  \draw [ultra thick, -<-=0.6] (-120:1.5) -- +(-120:1)
  node[shift={(-120:8pt)}] {$\beta^n$};

  \draw [thick, -<-=0.6] (120:1.5) -- +(120:1)
  node[shift={(120:6pt)}] {$-\bs h_1$};

  \draw [thick, -<-=0.6] (180:1.5) -- +(180:1)
  node[shift={(180:10pt)}] {$\bs h_N$};

  \node[shift={(150:8pt)}, xshift=-4pt] at (150:1.5) {$\Delta_{i^1} \alpha^1$};

  \node[shift={(90:-7pt)}] at (90:1.5) {$\alpha^1$};
  \node[shift={(30:-8pt)}] at (30:1.5) {$\alpha^2$};
  \node[shift={(-150:-15pt)}, yshift=-2pt] at (-150:1.5) {$\alpha^{n+1}$};
\end{tikzpicture}
\,.
\end{equation}
The difference operator $\Delta_i$ acts on internal momenta by
\begin{equation}
  \Delta_i\alpha = \alpha - \bs h_i \,.
\end{equation}
The function $F_{i^1}$ is given by
\begin{equation}
  F_{i^1}
  =
  \frac{\Gamma(N\bs q)}{\Gamma(\bs q)}
  \prod_{j^1 (\neq i^1)}
  \frac{\Gamma(\iu\bs \as^1_{j^1i^1})}{\Gamma(\bs q + \iu\bs \as^1_{j^1i^1})}
  \,,
\end{equation}
with
\begin{equation}
  Q = q\rho \,,
  \qquad
  \rho = \sum_{i=1}^{N-1} \omega_i \,.
\end{equation}

Next, we transport $V_{-\bs h_1}$ along $\gamma_\sigma$.  Graphically,
we move the external edge labeled $-\bs h_1$ clockwise.  Every time
the line passes another external edge we get a braiding factor:
\begin{equation}
\begin{aligned}
\begin{tikzpicture}[scale=0.5, baseline=(x.base)]
  \node (x) at (0,0) {\vphantom{x}};

  \draw [ultra thick, ->-=0.85] (0,0) circle [radius=1.5];

  \draw [ultra thick, -<-=0.6] (60:1.5) -- +(60:1)
  node[shift={(60:8pt)}] {$\beta^1$};

  \draw [ultra thick, -<-=0.6] (0:1.5) -- +(0:1)
  node[shift={(0:8pt)}] {$\beta^2$};

  \draw [ultra thick, dotted] (-40:2) arc (-40:-80:2);

  \draw [ultra thick, -<-=0.6] (-120:1.5) -- +(-120:1)
  node[shift={(-120:8pt)}] {$\beta^n$};

  \draw [thick, -<-=0.6] (120:1.5) -- +(120:1)
  node[shift={(120:6pt)}] {$-\bs h_1$};

  \draw [thick, -<-=0.6] (180:1.5) -- +(180:1)
  node[shift={(180:10pt)}] {$\bs h_N$};

  \node[shift={(150:8pt)}, xshift=-4pt] at (150:1.5) {$\Delta_{i^1} \alpha^1$};

  \node[shift={(90:-7pt)}] at (90:1.5) {$\alpha^1$};
  \node[shift={(30:-8pt)}] at (30:1.5) {$\alpha^2$};
  \node[shift={(-150:-15pt)}, yshift=-2pt] at (-150:1.5) {$\alpha^{n+1}$};
\end{tikzpicture}
&=
\sum_{i^2} B^{\sigma^2}_{i^1 i^2}
\begin{tikzpicture}[scale=0.5, baseline=(x.base)]
  \node (x) at (0,0) {\vphantom{x}};

  \draw [ultra thick, ->-=0.85] (0,0) circle [radius=1.5];

  \draw [ultra thick, -<-=0.6] (120:1.5) -- +(120:1)
  node[shift={(120:6pt)}] {$\beta^1$};

  \draw [ultra thick, -<-=0.6] (0:1.5) -- +(0:1)
  node[shift={(0:8pt)}] {$\beta^2$};

  \draw [ultra thick, dotted] (-40:2) arc (-40:-80:2);

  \draw [ultra thick, -<-=0.6] (-120:1.5) -- +(-120:1)
  node[shift={(-120:8pt)}] {$\beta^n$};

  \draw [thick, -<-=0.6] (60:1.5) -- +(60:1)
  node[shift={(60:6pt)}] {$-\bs h_1$};

  \draw [thick, -<-=0.6] (180:1.5) -- +(180:1)
  node[shift={(180:10pt)}] {$\bs h_N$};

  \node[shift={(150:8pt)}, xshift=-4pt] at (150:1.5) {$\Delta_{i^1} \alpha^1$};
  \node[shift={(90:8pt)}, xshift=2pt] at (90:1.5) {$\Delta_{i^2} \alpha^2$};

  \node[shift={(30:-8pt)}] at (30:1.5) {$\alpha^2$};
  \node[shift={(-150:-15pt)}, yshift=-2pt] at (-150:1.5) {$\alpha^{n+1}$};
\end{tikzpicture}
\\
&=
\sum_{i^2, \dotsc, i^{n+1}} \Biggl(\prod_{r=1}^n B^{\sigma^r}_{i^r i^{r+1}}\Biggr)
\begin{tikzpicture}[scale=0.5, baseline=(x.base)]
  \node (x) at (0,0) {\vphantom{x}};

  \draw [ultra thick, ->-=0.85] (0,0) circle [radius=1.5];

  \draw [ultra thick, -<-=0.6] (60:1.5) -- +(60:1)
  node[shift={(60:8pt)}] {$\beta^1$};

  \draw [ultra thick, -<-=0.6] (0:1.5) -- +(0:1)
  node[shift={(0:8pt)}] {$\beta^2$};

  \draw [ultra thick, dotted] (-40:2) arc (-40:-80:2);

  \draw [ultra thick, -<-=0.6] (-120:1.5) -- +(-120:1)
  node[shift={(-120:8pt)}] {$\beta^n$};

  \draw [thick, -<-=0.6] (120:1.5) -- +(120:1)
  node[shift={(120:6pt)}] {$\bs h_N$};

  \draw [thick, -<-=0.6] (180:1.5) -- +(180:1)
  node[shift={(180:12pt)}] {$-\bs h_1$};

  \node[shift={(90:8pt)}, xshift=2pt] at (90:1.5) {$\Delta_{i^1} \alpha^1$};
  \node[shift={(30:8pt)}, xshift=8pt] at (30:1.5) {$\Delta_{i^2} \alpha^2$};
  \node[shift={(-150:10pt)}, xshift=-16pt] at (-150:1.5)
  {$\Delta_{i^{n+1}} \alpha^{n+1}$};
  
  \node[shift={(150:-14pt)}, yshift=6pt] at (150:1.5) {$\alpha^{n+1}$};
\end{tikzpicture} \,.
\end{aligned}
\end{equation}
The function $B^{\sigma^r}_{i^r i^{r+1}}$ depends on the sign $\sigma^r$,
which specifies the direction of the braiding moves:
\begin{multline}
  B^{\sigma^r}_{i^r i^{r+1}}
  =
  e^{-\sigma^r \pi\bs(\as^r_{i^r} - \as^{r+1}_{i^{r+1}})}
  \prod_{j^r (\neq i^r)}
  \frac{\Gamma(\bs(q + \iu \as^r_{j^r i^r}))}
       {\Gamma(\bs(\frac12 q + \iu \as^r_{j^r} - \iu \as^{r+1}_{i^{r+1}} - \iu \ms^r))}
  \\
  \times     
  \prod_{j^{r+1} (\neq i^{r+1})}
  \frac{\Gamma(\iu\bs \as^{r+1}_{j^{r+1} i^{r+1}})}
       {\Gamma(\bs(\frac12 q + \iu \as^{r+1}_{j^{r+1}} - \iu \as^r_{i^r} + \iu \ms^r))}
  \,.
\end{multline}

Finally, we fuse $V_{-\bs h_1}$ and $V_{\bs h_N}$ and project the
result to the channel in which the intermediate state is the identity
operator:
\begin{equation}
\begin{tikzpicture}[scale=0.5, baseline=(x.base)]
  \node (x) at (0,0) {\vphantom{x}};

  \draw [ultra thick, ->-=0.85] (0,0) circle [radius=1.5];

  \draw [ultra thick, -<-=0.6] (60:1.5) -- +(60:1)
  node[shift={(60:8pt)}] {$\beta^1$};

  \draw [ultra thick, -<-=0.6] (0:1.5) -- +(0:1)
  node[shift={(0:8pt)}] {$\beta^2$};

  \draw [ultra thick, dotted] (-40:2) arc (-40:-80:2);

  \draw [ultra thick, -<-=0.6] (-120:1.5) -- +(-120:1)
  node[shift={(-120:8pt)}] {$\beta^n$};

  \draw [thick, -<-=0.6] (120:1.5) -- +(120:1)
  node[shift={(120:6pt)}] {$\bs h_N$};

  \draw [thick, -<-=0.6] (180:1.5) -- +(180:1)
  node[shift={(180:12pt)}] {$-\bs h_1$};

  \node[shift={(90:8pt)}, xshift=2pt] at (90:1.5) {$\Delta_{i^1} \alpha^1$};
  \node[shift={(30:8pt)}, xshift=8pt] at (30:1.5) {$\Delta_{i^2} \alpha^2$};
  \node[shift={(-150:10pt)}, xshift=-16pt] at (-150:1.5)
  {$\Delta_{i^{n+1}} \alpha^{n+1}$};
  
  \node[shift={(150:-14pt)}, yshift=6pt] at (150:1.5) {$\alpha^{n+1}$};
\end{tikzpicture}
\longrightarrow
\frac{\sin(\pi \bs q)}{\sin(\pi N\bs q)} F_{i^1}^{-1}
\begin{tikzpicture}[scale=0.5, baseline=(x.base)]
  \node (x) at (0,0) {\vphantom{x}};
  
  \draw [ultra thick, ->-=0.85] (0,0) circle [radius=1.5];

  \draw [ultra thick, -<-=0.6] (60:1.5) -- +(60:1)
  node[shift={(60:8pt)}] {$\beta^1$};

  \draw [ultra thick, -<-=0.6] (0:1.5) -- +(0:1)
  node[shift={(0:8pt)}] {$\beta^2$};

  \draw [ultra thick, dotted] (-40:2) arc (-40:-80:2);

  \draw [ultra thick, -<-=0.6] (-120:1.5) -- +(-120:1)
  node[shift={(-120:8pt)}] {$\beta^n$};

  \draw [thick, densely dashed, -<-=0.6] (150:1.5) --
  node[shift={(60:6pt)}] {$0$} +(150:1);

  \draw [thick, -<-=0.6] (150:2.5) -- +(90:1)
  node[shift={(90:6pt)}] {$\bs h_N$};

  \draw [thick, -<-=0.6] (150:2.5) -- +(210:1)
  node[shift={(210:8pt)}] {$-\bs h_1$};

  \node[shift={(105:8pt)}, xshift=4pt] at (105:1.5)
  {$\Delta_{i^1} \alpha^1$};
  \node[shift={(30:8pt)}, xshift=8pt] at (30:1.5)
  {$\Delta_{i^2} \alpha^2$};
  \node[shift={(-165:24pt)}, xshift=0pt] at (-165:1.5)
  {$\Delta_{i^{n+1}} \alpha^{n+1}$};
\end{tikzpicture}
\,.
\end{equation}
Note that the right-hand side vanishes unless $i^{n+1} = i^1$ since
$\alpha^{n+1} = \alpha^1$.

Thus, dropping the overall factor $\sin(\pi \bs q)/\sin(\pi N\bs q)$, we
find that the Verlinde operator is the difference operator
\begin{equation}
  \Phi_{-\bs h_1}(\gamma_\sigma)
  =
  \sum_{i^1, \dotsc, i^n}
  \biggl(\prod_r B^{\sigma^r}_{i^r i^{r+1}}\biggr)
  \Delta_{\{i^1,\dotsc,i^n\}}  \,,
\end{equation}
where $i^{n+1} = i^1$ and
\begin{equation}
  \Delta_{\{i^1,\dotsc,i^n\}}
  =
  \prod_r \Delta^r_{i^r} \,.
\end{equation}

Before we compare the Verlinde operator with the transfer matrix, we
must perform a change of basis and find the modified Verlinde
operator~\eqref{eq:c-verlinde}.  For the correlation function at hand,
the product of three-point function factors is
\begin{equation}
  \CC(\alpha; \beta)
  =
  \prod_r
  \frac{\prod_{i<j} \Upsilon(\iu \as^r_{ij}) \Upsilon(-\iu \as^{r+1}_{ij})}
       {\prod_{i,j} \Upsilon(\frac12 q + \iu \ms^r - \iu \as^r_i + \iu \as^{r+1}_j)}
  \,.
\end{equation}
The precise definition of the function $\Upsilon$ is not important for
us; we just need to know that it satisfies the identity
\begin{equation}
  \frac{\Upsilon(x + \bs)}{\Upsilon(x)}
  = \frac{\Gamma(\bs x)}{\Gamma(1 - \bs x)} \bs^{1 - 2\bs x} \,,
\end{equation}
where $\Gamma$ is the gamma function.

Let us calculate
$\CC(\alpha; \beta) \Delta_{\{i^1,\dotsc,i^n\}} \CC(\alpha;
\beta)^{-1}$.  The only nontrivial contributions come from the
$\Upsilon$-factors in which either of $i$ or $j$ (but not both) in
$\as^r_{ij}$ is equal to $i^r$:
\begin{equation}
  \begin{split}
    & \CC(\alpha; \beta) \Delta_{\{i^1,\dotsc,i^n\}} \CC(\alpha; \beta)^{-1}
    \\
    &=
    \prod_r
    \prod_{(i^r<)j}
    \frac{\Upsilon(\iu \as^r_{i^r j})}{\Upsilon(\iu \as^r_{i^r j} - \bs)}
    \prod_{i(<i^r)}
    \frac{\Upsilon(\iu \as^r_{ii^r})}{\Upsilon(\iu \as^r_{ii^r} + \bs)}
    \\ & \quad
    \times
    \prod_{(i^{r+1}<)j}
    \frac{\Upsilon(-\iu \as^{r+1}_{i^{r+1} j})}
         {\Upsilon(-\iu \as^{r+1}_{i^{r+1} j} + \bs)}
    \prod_{i(<i^{r+1})}
    \frac{\Upsilon(-\iu \as^{r+1}_{ii^{r+1}})}
         {\Upsilon(-\iu \as^{r+1}_{ii^{r+1}} - \bs)}
    \\ & \quad
    \times
    \prod_{i(\neq i^r)}
    \frac{\Upsilon(\frac12 q + \iu \ms^r - \iu \as^r_{i} + \iu \as^{r+1}_{i^{r+1}} - \bs)}
         {\Upsilon(\frac12 q + \iu \ms^r - \iu \as^r_{i} + \iu \as^{r+1}_{i^{r+1}})}
    \prod_{j(\neq i^{r+1})}
    \frac{\Upsilon(\frac12 q + \iu \ms^r - \iu \as^r_{i^r} + \iu \as^{r+1}_{j} + \bs)}
         {\Upsilon(\frac12 q + \iu \ms^r - \iu \as^r_{i^r} + \iu \as^{r+1}_{j})}
 \,.
  \end{split}
\end{equation}
Combining the first two lines and using the aforementioned identity,
we can rewrite this quantity as
\begin{equation}
  \begin{split}
    &\CC(\alpha; \beta) \Delta_{\{i^1,\dotsc,i^n\}} \CC(\alpha; \beta)^{-1}
    \\
    &=
    \prod_r
    \prod_{j^r (\neq i^r)}
    \frac{\Gamma(1 - \bs(q + \iu \as^r_{j^r i^r}))
        \Gamma(\bs(\frac12 q + \iu \as^r_{j^r} - \iu \as^{r+1}_{i^{r+1}} - \iu \ms^r))}
      {\Gamma(\bs(q + \iu \as^r_{j^r i^r}))
       \Gamma(1 - \bs(\frac12 q + \iu \as^r_{j^r} - \iu \as^{r+1}_{i^{r+1}} - \iu \ms^r))}
    \\
    & \quad
    \times
    \prod_{j^{r+1}(\neq i^{r+1})}
    \frac{\Gamma(1 - \iu\bs \as^{r+1}_{j^{r+1} i^{r+1}})
      \Gamma(\bs(\frac12 q + \iu \as^{r+1}_{j^{r+1}} - \iu \as^r_{i^r} + \iu \ms^r))}
    {\Gamma(\iu\bs\as^{r+1}_{j^{r+1} i^{r+1}})
     \Gamma(1 - \bs(\frac12 q + \iu \as^{r+1}_{j^{r+1}} - \iu \as^r_{i^r} + \iu \ms^r))}
    \,.
  \end{split}
\end{equation}

Plugging this expression into the formula for the modified Verlinde
operator, we see that the various factors of gamma functions combine
nicely into sine functions via Euler's reflection formula
\begin{equation}
  \Gamma(x) \Gamma(1-x) = \frac{\pi}{\sin(\pi x)} \,.
\end{equation}
The final result is
\begin{multline}
  \CL_{-\bs h_1}(\gamma_\sigma)
  =
  \sum_{i^1, \dotsc, i^n}
  \Biggl(
  \prod_r
  e^{-\sigma^r \pi\bs(\as^r_{i^r} - \as^{r+1}_{i^{r+1}})}
  \prod_{j^r (\neq i^r)}
  \Biggl(
  \frac{\sin\pi\bs(\frac12 q + \iu \as^r_{j^r} - \iu \as^{r+1}_{i^{r+1}} - \iu \ms^r)}
       {\sin\pi\bs(q + \iu \as^r_{j^r i^r})}
  \Biggr)^{\frac12}
  \\
  \times
  \prod_{j^{r+1} (\neq i^{r+1})}
  \Biggl(
  \frac{\sin\pi\bs(\frac12 q + \iu \as^{r+1}_{j^{r+1}} - \iu \as^r_{i^r} + \iu \ms^r)}
       {\sin\pi \iu\bs \as^{r+1}_{j^{r+1} i^{r+1}}}
  \Biggr)^{\frac12}
  \Biggr)
  \Delta_{\{i^1,\dotsc,i^n\}} \,.
\end{multline}

The above expression can be written in terms of the
functions~\eqref{eq:ell} as
\begin{multline}
  \CL_{-\bs h_1}(\gamma_\sigma)
  \\
  =
  \sum_{i^1, \dotsc, i^n}
  \Delta_{\{i^1,\dotsc,i^n\}}^{\frac12}
  \biggl(
  \prod_r
  \ell_{\iu\bs \ms^r + \frac12}(\iu\bs \as^r, \iu\bs \as^{r+1})_{i^r}^{i^{r+1}}
  e^{\sigma^r \pi\iu(\iu\bs \as^r_{i^r} - \iu\bs \as^{r+1}_{i^{r+1}})}
  \biggr)
  \Delta_{\{i^1,\dotsc,i^n\}}^{\frac12} \,.
\end{multline}
Comparing this expression with the Wigner transform~\eqref{eq:T-trig}
of the trigonometric transfer matrix, we deduce that the modified
Verlinde operator coincides with the transfer matrix,
\begin{equation}
  \CL_{-\bs h_1}(\gamma_\sigma) = \CT_{\sigma,m} \,,
\end{equation}
under the identification
\begin{equation}
  \eps = \bs^2 \,,
  \qquad
  a^r= \iu\bs\as^r \,,
  \qquad
  m^r = \iu\bs\ms^r + \frac12 \,.
\end{equation}

It has been proposed in~\cite{Ito:2011ea} that precisely under this
identification of parameters, a modified Verlinde operator in Toda
theory corresponding to a Wilson--'t Hooft line in the AGT-dual theory
on $S^4_\bs$ reproduces the Weyl quantization of the same Wilson--'t
Hooft line in the same theory, but placed in the spacetime
$S^1 \times_\eps \R^2 \times \R$.  Therefore, we again reach the
conclusion that the vev of the Wilson--'t Hooft line with
charge~\eqref{eq:B-circular} and~\eqref{eq:E-circular} are equal to
the Wigner transform of the trigonometric transfer
matrix~\eqref{eq:T-trig}.

\section{Brane realization}
\label{sec:branes}

The AGT correspondence between Wilson--'t Hooft lines and Verlinde
operators, which we exploited in section~\ref{sec:Toda}, can be
realized in terms of branes in string theory.  String dualities relate
the brane configuration for the AGT correspondence to another
configuration that realizes four-dimensional Chern--Simons theory, and
in the latter setup the emergence of quantum integrability can be seen
more transparently.  Another chain of dualities relate these setups to
the one studied in~\cite{Maruyoshi:2016caf, Yagi:2017hmj}, which
provided the initial motivation for the present work.  In this last
section we discuss these brane constructions.

As explained in section~\ref{sec:Toda}, the field theoretic origin of
the AGT correspondence is six-dimensional $\CN = (2,0)$ superconformal
field theory, which in our context is of type $A_{N-1}$ and
compactified on an $n$-punctured torus $C_{1,n}$.  This theory
describes the low-energy dynamics of a stack of $N$ M5-branes (modulo
the center-of-mass degrees of freedom), intersected by $n$ M5-branes.

Consider M-theory in the eleven-dimensional spacetime
\begin{equation}
  M_{11}
  =
  \R_0 \times \R^2_{12} \times_\eps S^1_3 \times_{-\eps} \R^2_{45}
  \times S^1_6 \times \R_7 \times \R_8 \times \R_9 \times S^1_{10} \,.
\end{equation}
(The subscripts indicate the directions in which the spaces extend.)
We put $N$ M5-branes $\mathrm{M5}_i$, $i = 1$, $\dotsc$, $N$, on
\begin{equation}
  M_{\mathrm{M5}_i} =
  \R_0 \times \R^2_{12} \times_\eps S^1_3 \times \{0\}
  \times S^1_6 \times \{0\} \times \{0\} \times \{0\} \times S^1_{10} \,.
\end{equation}
They realize $\CN = (2,0)$ superconformal field theory on
$\R_0 \times \R^2_{12} \times_\eps S^1_3 \times C_1$, with
\begin{equation}
  C_1 = S^1_6 \times S^1_{10} \,.
\end{equation}
Further, we introduce $n$ M5-branes $\mathrm{M5}^r$, $r = 1$,
$\dotsc$, $n$, with worldvolumes
\begin{equation}
  M_{\mathrm{M5}^r} =
  \R_0 \times \R^2_{12} \times_\eps S^1_3 \times \{0\}
  \times \{l^r\} \times \{0\} \times \R_8 \times \R_9 \times \{\theta^r\} \,.
\end{equation}
These M5-branes create codimension-two defects in the six-dimensional
theory, located at $n$ points $(l^r,\theta^r)$ on $C_1$, making $C_1$
an $n$-punctured torus $C_{1,n}$.

The two sets of M5-branes share a four-dimensional part of the
spacetime, $\R_0 \times \R^2_{12} \times_\eps S^1_3$, and on this
four-dimensional spacetime we get an $\CN = 2$ supersymmetric gauge
theory with gauge group $G = \SU(N)^n$, described by the circular
quiver with $n$ nodes.  (More precisely, the gauge group is
$\SU(N)^n \times \U(1)$ but the $\U(1)$ factor is associated with the
center-of-mass and decoupled from the rest of the theory.)  In fact,
reduction on $S^1_{10}$ turns $\mathrm{M5}_i$ into D4-branes
$\mathrm{D4}_i$ on
\begin{equation}
  M_{\mathrm{D4}_i} =
  \R_0 \times \R^2_{12} \times_\eps S^1_3 \times \{0\}
  \times S^1_6 \times \{0\} \times \{0\} \times \{0\}
\end{equation}
and $\mathrm{M5}^r$ into NS5-branes $\mathrm{NS5}^r$ on
\begin{equation}
  M_{\mathrm{NS5}^r} =
  \R_0 \times \R^2_{12} \times_\eps S^1_3 \times \{0\}
  \times \{l^r\} \times \{0\} \times \R_8 \times \R_9 \,,
\end{equation}
and the above brane configuration becomes the well-known D4--NS5 brane
configuration for the circular quiver theory~\cite{Witten:1997sc}.
The difference $l^{r+1} - l^r$ in the $x^6$-coordinate between
$\mathrm{NS5}^{r+1}$ and $\mathrm{NS5}^r$ is inversely proportional to
the square of the gauge coupling for the $r$th gauge group, whereas
the difference $\theta^{r+1} - \theta^r$ in the $x^{10}$-coordinate is
the theta-angle for the $r$th gauge group.%
\footnote{To realize nonzero values for the parameters $a^r$ and
  $b^r$, we break each D4-brane $\mathrm{D4}_i$ into $n$ segments
  $\mathrm{D4}_i^r$ suspended between neighboring NS5-branes and allow
  these segments to be located anywhere on $\R_8 \times \R_9$.  Then,
  $a^r_i$ is a complex linear combination of the $x^9$-coordinate of
  $\mathrm{D4}_i^r$ and the background holonomy of the $\U(1)$ gauge
  field on $\mathrm{D4}_i^r$ around $S^1_3$.  The definition of
  $b_i^r$ is similar, but involves both the $x^8$- and
  $x^9$-coordinates as well as a chemical potential for the magnetic
  charge at infinity which does not have a simple interpretation in
  this brane system.}

A Wilson--'t Hooft line in this four-dimensional theory is realized by
an M2-brane on
\begin{equation}
  M_{\mathrm{M2}} = 
  \{t_0\} \times \{0\} \times S^1_3 \times \{0\}
  \times S^1_6 \times \{0\} \times \R^{\geq0}_8
  \times \{x_0\} \times \{\theta_0\} \,,
\end{equation}
where $\R^{\geq0}_8$ is the nonnegative part of $\R_8$.  Upon reduction on $S^1_{10}$, this M2-brane becomes a D2-brane on 
\begin{equation}
  M_{\mathrm{D2}} = 
  \{t_0\} \times \{0\} \times S^1_3 \times \{0\}
  \times S^1_6 \times \{0\} \times \R^{\geq0}_8
  \times \{x_0\}
\end{equation}
and creates a Wilson--'t Hooft line of the type considered in
section~\ref{sec:gauge}.  It corresponds to a Verlinde operator in
Toda theory on $C_{1,n}$, constructed from a vertex operator
transported along the path $S^1_6 \times \{\theta_0\}$.  We will
explain in a moment how to get the other relevant Verlinde operators.

To understand the relation to quantum integrable systems, let us
compactify $\R_9$ to a circle $S^1_9$ of radius $R_9$.  By doing so,
we are uplifting the four-dimensional gauge theory to a
five-dimensional one, compactified on a circle.  Indeed, by T-duality
on $S^1_9$ we get D5-branes $\check{\mathrm{D5}}_i$, NS5-branes
$\check{\mathrm{NS5}}^r$ and a D3-brane $\check{\mathrm{D3}}$ with
worldvolumes
\begin{align}
  \label{ea:M-D5}
  M_{\check{\mathrm{D5}}_i}
  &=
  \R_0 \times \R^2_{12} \times_\eps S^1_3 \times \{0\}
  \times S^1_6 \times \{0\} \times \{0\} \times \check S^1_9 \,,
  \\
  \label{ea:M-NS5}
  M_{\check{\mathrm{NS5}}^r}
  &=
  \R_0 \times \R^2_{12} \times_\eps S^1_3 \times \{0\}
  \times \{l^r\} \times \{0\} \times \R_8 \times \check S^1_9 \,,
  \\
  \label{ea:M-D3}
  M_{\check{\mathrm{D3}}}
  &= 
  \{t_0\} \times \{0\} \times S^1_3 \times \{0\}
  \times S^1_6 \times \{0\} \times \R^{\geq0}_8
  \times \check S^1_9 \,.
\end{align}
The D5- and NS5-branes intersect along
$\R_0 \times \R^2_{12} \times_\eps S^1_3 \times \check S^1_9$, where a
five-dimensional circular quiver theory arises.  Recall that the radius
$\check{R}_9$ of the dual circle $\check{S}^1_9$ is inversely
proportional to the original radius, $\check{R}_9 = \alpha'/R_9$.

Going back to the M-theory setup, we reduce it on $S^1_3$ and apply
T-duality on $S^1_9$.  Then, $\mathrm{M5}_i$ become D5-branes
$\widetilde{\mathrm{D5}}_i$ on
\begin{equation}
  M_{\widetilde{\mathrm{D5}_i}}
  =
  \R_0 \times \R^2_{12} \times \{0\} \times S^1_6
  \times \{0\} \times \{0\} \times \check{S}^1_9 \times S^1_{10} \,,
\end{equation}
$\mathrm{M5}^r$ become D3-branes $\widetilde{\mathrm{D3}}^r$ on
\begin{equation}
  M_{\widetilde{\mathrm{D3}}^r}
  =
  \R_0 \times \R^2_{12} \times \{0\} \times \{l^r\}
  \times \{0\} \times \R_8 \times \{\phi^r\} \times \{\theta^r\} \,,
\end{equation}
and $\mathrm{M2}$ becomes a fundamental string $\widetilde{\mathrm{F1}}$ on
\begin{equation}
  M_{\widetilde{\mathrm{F1}}}
  =
  \{t_0\} \times \{0\} \times \{0\}
  \times S^1_6 \times \{0\} \times \R^{\geq0}_8
  \times \{\phi_0\} \times \{\theta_0\} \,.
\end{equation}

The $N$ D5-branes $\widetilde{\mathrm{D5}}_i$ support $\CN = (1,1)$
super Yang--Mills theory with gauge group $\SU(N)$ on
$\R_0 \times \R^2_{12} \times S^1_6 \times \check{S}^1_9 \times
S^1_{10}$.  This theory is, however, deformed because the product
between $S^1_3$ and $\R^2_{12} \times \R^2_{45}$ was twisted, which
induces nontrivial background fields after reduction on $S^1_3$.  This
deformation is of the type studied in~\cite{Yagi:2014toa} and the same
as the one applied to the D5-brane theory in~\cite{Costello:2018txb}.%
\footnote{The equivalence to the construction
  of~\cite{Costello:2018txb} can be seen as follows.  Let us reduce
  the M-theory setup instead on $S^1_{10}$ and perform T-duality on
  $S^1_9$.  Then, we obtain a type IIB setup whose geometry contains a
  twisted product between $\R^2_{12} \times \R^2_{45}$ and the torus
  $S^1_3 \times \check S^1_9$.  This is the starting point of the
  construction of~\cite{Costello:2018txb}.  The string background
  proposed in~\cite{Costello:2018txb} is achieved by a sequence of
  dualities applied to this setup: T-duality on
  $S^1_3 \times \check S^1_9$, then S-duality, and finally T-duality
  on the dual torus $\check S^1_3 \times S^1_9$.  The first T-duality
  realizes the $\Omega$-deformation~\cite{Hellerman:2011mv}, hence we
  are considering the S-dual $\Omega$-deformation here.  The point is
  that S-duality is geometrized via M-theory as a ``9-11 flip.''  In
  the above construction, the S-duality can be replaced by T-duality
  on $\check S^1_3$, a lift to M-theory and reduction on $S^1_3$, and
  T-duality on the M-theory circle $S^1_{10}$.  The configuration
  right after the lift to M-theory is nothing but our original
  M-theory setup.  The subsequent steps in the 9-11 flip differ from
  what we did to that setup only in the choice of the circle on which
  we perform T-duality, $S^1_{10}$ now whereas $S^1_9$ before.
  However, we are still supposed to perform T-duality on
  $S^1_9 \times \check S^1_{10}$ as the last step in the construction
  of~\cite{Costello:2018txb}.  (Note that the 9-11 flip has exchanged
  $S^1_3$ and $S^1_{10}$.)  This fixes the discrepancy.}
If we consider the reduction to four-dimensional $\CN = 4$ super
Yang--Mills theory, this deformation descends to the S-dual of the
$\Omega$-deformation~\cite{Nekrasov:2002qd}.

In the sector in which the relevant supersymmetry is preserved, the
deformation leads to localization of the path integral.  As a result,
this sector of the deformed theory is equivalent to a bosonic theory
which, roughly speaking, may be understood as living at the origin of
$\R^2_{12}$.  This theory turns out to be a four-dimensional variant
of Chern--Simons theory~\cite{Costello:2018txb}, with Planck constant
$\hbar \propto \eps$.  Four-dimensional Chern--Simons theory, here
placed on $\R_0 \times S^1_6 \times \check{S}^1_9 \times S^1_{10}$,
depends topologically on the cylinder
\begin{equation}
  \Sigma = \R_0 \times S^1_6
\end{equation}
and holomorphically on the torus
\begin{equation}
  E = \check{S}^1_9 \times S^1_{10} \,.
\end{equation}

The D3-branes $\widetilde{\mathrm{D3}}^r$ create line defects
extending in the longitudinal direction of $\Sigma$ and located at the
points
\begin{equation}
  w^r = \phi^r + \iu\theta^r
\end{equation}
on $E$.  The fundamental string $\widetilde{\mathrm{F1}}$, on the
other hand, creates a Wilson line in the vector representation that
winds around the circumferential direction and is located at
\begin{equation}
  z_0 = \phi_0 + \iu\theta_0
\end{equation}
on $E$.  Thus, on the cylinder $\Sigma$, we have the same situation as
in figure~\ref{fig:spin-chain-T}, in which a quantum spin chain was
described in terms of lines on a cylinder.

Indeed, a quantum integrable system emerges from such a configuration
of line operators in four-dimensional Chern--Simons
theory~\cite{Costello:2013zra}.  The Hilbert space of the integrable
system is the space of states of the field theory on a time slice
(where the $x^0$-coordinate is constant) intersected by line operators
extending in the time direction.  On this Hilbert space act transfer
matrices, which are Wilson lines in the $x^6$-direction.  The
integrability is a consequence of the topological--holomorphic nature
of the theory: by the topological invariance on $\Sigma$, one can
slide line operators winding around the cylinder continuously along
the longitudinal direction; and if two such line operators are located
at different points on $E$, one can move them past each other without
encountering a phase transition, thereby establishing the
commutativity of transfer matrices.

It was argued in~\cite{Costello:2018txb}, based on the earlier
work~\cite{Maruyoshi:2016caf,Yagi:2017hmj}, that a crossing of line
defects created by a D3-brane and a fundamental string produces the
elliptic L-operator~\eqref{eq:L-ell} with $z = z_0$, $w = w^r$ and
$y = 0$%
\footnote{More generally, $\mathrm{D3}^r$ can be split into two
  semi-infinite D3-branes $\mathrm{D3}^r_+$ and $\mathrm{D3}^r_-$,
  each ending on the stack of D5-branes at $x^8 = 0$.  The parameter
  $y$ is given by the separation of these two halves in $E$.  In the
  five-dimensional circular quiver theory, the separation is
  proportional to the complex mass parameter $m^r$ for the
  bifundamental hypermultiplet charged under the $r$th and $(r+1)$th
  gauge groups.}
(up to shifts by constants).  The parameter $\tau$ is the modulus of
$E$:
\begin{equation}
  \tau = \iu \frac{R_{10}}{\check{R}_9} \,.
\end{equation}

Now, take the limit $\check{R}_9 \to 0$, in which $\check{S}^1_9$
shrinks to a point, $S^1_9$ decompactifies, and the five-dimensional
circular quiver theory reduces to the four-dimensional one.  This is
the trigonometric limit $\tau \to \iu\infty$, so we conclude that the
transfer matrix constructed from the trigonometric L-operator arises
from a Wilson--'t Hooft line in the four-dimensional circular quiver
theory.

In the previous sections we studied the transfer matrix
$\CT_{\sigma,m}$ associated with the cycle $\gamma_\sigma$ in
$C_{1,n}$ specified by an $n$-tuples of signs $\sigma$.  The
Wilson--'t Hooft line considered above corresponds to a specific
choice of $\sigma$.  Those corresponding to the other choices can also
be constructed in a similar manner, but the construction is a little
more subtle.  Let us explain how this construction works from the
point of view of four-dimensional Chern--Simons theory.

\begin{figure}
  \centering
  \subfloat[]{
    \begin{tikzpicture}[scale=0.4, baseline=(x.base)]
      \node (x) at (0,0) {\vphantom{x}};
      
      \draw[fill=brown!30] (0,0) rectangle (10,6);
      
      \draw[thick, ->] (0,3) -- (2.5,3) -- (3.5,5) -- (4.5,3) --
      (6.5,3) -- (7.5,1) -- (8.5,3) -- (10,3);
      
      \draw[thick, double, ->] (3,0) -- +(90:6);
      \draw[thick, double, ->] (7,0) -- +(90:6);
      
      \node[right] at (0,5.3) {$\Sigma$};
      
      \draw[->] (-1,-1) -- +(90:1.5) node[above] {$x^0$};
      \draw[->] (-1,-1) -- +(0:1.5) node[right] {$x^6$};

      \node at (-1,3) {$\phi_0$};
      \node at (3,-1) {$l^r$};
      \node at (7,-1) {$l^{r+1}$};
    \end{tikzpicture}
  }
  \qquad
  \subfloat[]{
    \begin{tikzpicture}[scale=0.4, baseline=(x.base)]
      \node (x) at (0,0) {\vphantom{x}};
      
      \draw[fill={rgb,255:red,208; green,181; blue,155}] (0,0) rectangle (10,6);
      
      \node[black!70] (p1) at (3,3) {$\bullet$};
      \node[black!70] (p2) at (7,3) {$\bullet$};
        
      \draw[thick, ->] (0,3) -- (2.5,3) -- (2.5,1) -- (3.5,1) -- (3.5,5) -- (4.5,5) -- (4.5,3) -- (6.5,3) -- (6.5,5) -- (7.5,5) -- (7.5,1) -- (8.5,1) -- (8.5,3) -- (10,3);
      
      \node[right] at (0,5.3) {$C_{1,n}$};

      \draw[->] (-1,-1) -- +(90:1.5) node[above] {$x^{10}$};
      \draw[->] (-1,-1) -- +(0:1.5) node[right] {$x^6$};

      \node at (-1,3) {$\theta_0$};
      \node at (3,-1) {$l^r$};
      \node at (7,-1) {$l^{r+1}$};
    \end{tikzpicture}
  }
  \caption{(a) A path in $\Sigma$ bending near double lines.  (b) The
    corresponding path in $C_{1,n}$ detours around the punctures.}
  \label{fig:path-4dCS}
\end{figure}
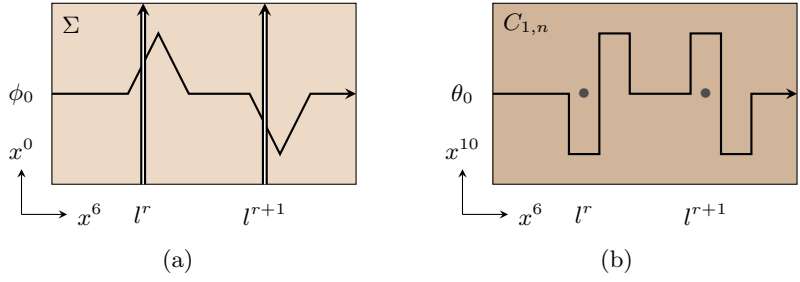

For simplicity, let us set all $\theta^r = \theta_0$.  (Since
$\CT_{\sigma,m}$ is independent of the spectral parameters $z_0$ and
$w^r$, we do not lose anything by this specialization.)  According to
the analysis of~\cite{Costello:2017dso}, framing anomaly requires that
if a Wilson line curves by an angle $\varphi$, its coordinate on
$S^1_{10}$ must be shifted by $-\eps N\varphi/2\pi$.  We can make use
of this property to get a Wilson line supported on the cycle
$\gamma_\sigma$: fix a small value $\varphi_0$ and let the Wilson line
bends by the angle $\sigma^r\varphi_0$ right before it crosses the
$r$th double line, as illustrated in figure~\ref{fig:path-4dCS}.

The trigonometric limit $\check{R}_9 \to 0$ is equivalent to the limit
$R_{10} \to \infty$.  In this limit, $E$ is elongated by an infinite
factor in the $x^{10}$-direction and the solid line is located at
$z = \phi_0 - \sigma^r \iu\infty$ when it crosses the $r$th double
line.  This is precisely the limit that appears in the definitions of
the fundamental L-operators~\eqref{eq:fund-L}, from which
$\CT_{\sigma,m}$ is constructed.

Let us summarize the relation between the circular quiver theory and
four-dimensional Chern--Simons theory.  The circular quiver theory
with gauge group $\SU(N)^n$ arises from six-dimensional $\CN = (2,0)$
superconformal field theory of type $A_{N-1}$ compactified on an
$n$-punctured torus, which is realized by a stack of $N$ M5-branes
intersected by $n$ M5-branes.  If we place this brane system in a
twisted geometry, then by string dualities it is mapped to $N$
D5-branes and $n$ D3-branes in a particular string background.  The
D5-branes support six-dimensional $\CN = (1,1)$ super Yang--Mills
theory with gauge group $\SU(N)$, and this background deforms it.  It
is a topological--holomorphic sector of this deformed theory that is
described by four-dimensional Chern--Simons theory.  The D3-branes
become line operators in four-dimensional Chern--Simons theory and
prepare the Hilbert space of the integrable system.  Finally, we can
include M2-branes that create Wilson--'t Hooft lines in the circular
quiver theory.  They are mapped to fundamental strings in the dual
frame and appear as Wilson lines in four-dimensional Chern--Simons
theory.  These Wilson lines act on the Hilbert space of the integrable
system by transfer matrices.

Lastly, we remark that the D5--NS5--D3 brane
system~\eqref{ea:M-D5}--\eqref{ea:M-D3} is another interesting duality
frame.  It is actually possible to introduce an additional set of
NS5-branes so that the 5-brane system realizes a four-dimensional
$\CN = 1$ supersymmetric gauge theory on
$\R^2_{12} \times_\eps S^1_3 \times \check S^1_9$.  The D3-brane
creates a surface defect in this theory.  As expected, it acts on the
partition function of the theory as an elliptic transfer
matrix~\cite{Maruyoshi:2016caf,Yagi:2017hmj}.

\acknowledgments

We would like to thank Takuya Okuda, Masato Taki and Yutaka Yoshida
for helpful discussions.  KM is grateful to Perimeter Institute for
hospitality.  The work of KM is supported in part by JSPS KAKENHI
Grant Number JP17K14296 and JP20K03935.  TO is supported in part by
JSPS KAKENHI Grant Number JP20J10126 and by RIKEN Junior Research
Associate Program.  The research of JY is supported by the Perimeter
Institute for Theoretical Physics.  Research at Perimeter Institute is
supported by the Government of Canada through the Department of
Innovation, Science and Economic Development and by the Province of
Ontario through the Ministry of Research, Innovation and Science.

\providecommand{\href}[2]{#2}\begingroup\raggedright\endgroup


\begin{thebibliography}{10}

\bibitem{Bullimore:2014nla}
M.~Bullimore, M.~Fluder, L.~Hollands and P.~Richmond, \emph{The superconformal
  index and an elliptic algebra of surface defects},
  \href{http://dx.doi.org/10.1007/JHEP10(2014)062}{\emph{JHEP} {\bf 10} (2014)
  062} [\href{http://arxiv.org/abs/1401.3379}{{\tt 1401.3379}}].

\bibitem{Maruyoshi:2016caf}
K.~Maruyoshi and J.~Yagi, \emph{Surface defects as transfer matrices},
  \href{http://dx.doi.org/10.1093/ptep/ptw151}{\emph{Prog. Theor. Exp.
  Phys.} (2016) 113B01, 52} [\href{http://arxiv.org/abs/1606.01041}{{\tt
  1606.01041}}].

\bibitem{Yagi:2017hmj}
J.~Yagi, \emph{Surface defects and elliptic quantum groups},
  \href{http://dx.doi.org/10.1007/JHEP06(2017)013}{\emph{JHEP} (2017) 013} [\href{http://arxiv.org/abs/1701.05562}{{\tt
  1701.05562}}].

\bibitem{Kapustin:2005py}
A.~Kapustin, \emph{Wilson--'t {H}ooft operators in four-dimensional gauge
  theories and {$S$}-duality},
  \href{http://dx.doi.org/10.1103/PhysRevD.74.025005}{\emph{Phys. Rev. D (3)}
  {\bf 74} (2006) 025005, 14}.

\bibitem{MR1463830}
K.~Hasegawa, \emph{Ruijsenaars' commuting difference operators as commuting
  transfer matrices},
  \href{http://dx.doi.org/10.1007/s002200050137}{\emph{Comm. Math. Phys.} {\bf
  187} (1997) 289--325}.

\bibitem{Baxter:1972wf}
R.~J. Baxter, \emph{{Eight-vertex model in lattice statistics and
  one-dimensional anisotropic Heisenberg chain. II\@. Equivalence to a
  generalized ice-type lattice model}},
  \href{http://dx.doi.org/10.1016/0003-4916(73)90440-5}{\emph{Ann. Phys.} {\bf
  76} (1973) 25--47}.

\bibitem{MR908997}
M.~Jimbo, T.~Miwa and M.~Okado, \emph{Solvable lattice models whose states are
  dominant integral weights of {$A^{(1)}_{n-1}$}},
  \href{http://dx.doi.org/10.1007/BF00420302}{\emph{Lett. Math. Phys.} {\bf 14}
  (1987) 123--131}.

\bibitem{Jimbo:1987mu}
M.~Jimbo, T.~Miwa and M.~Okado, \emph{Local state probabilities of solvable
  lattice models: an {$A^{(1)}_{n-1}$} family},
  \href{http://dx.doi.org/10.1016/0550-3213(88)90587-1}{\emph{Nucl. Phys. B}
  {\bf 300} (1988) 74--108}.

\bibitem{Pestun:2016zxk}
V.~Pestun, M.~Zabzine, F.~Benini and et~al., \emph{Localization techniques in
  quantum field theories},
  \href{http://dx.doi.org/10.1088/1751-8121/aa63c1}{\emph{J. Phys. A} {\bf 50}
  (2017) 440301} [\href{http://arxiv.org/abs/1608.02952}{{\tt 1608.02952}}].

\bibitem{Ito:2011ea}
Y.~Ito, T.~Okuda and M.~Taki, \emph{Line operators on {$S^1 \times
    \mathbb{R}^3$} and quantization of the {H}itchin moduli space},
\href{http://dx.doi.org/10.1007/JHEP04(2012)010}{\emph{JHEP}
  {\bf 04} (2012) 010} [\href{http://arxiv.org/abs/1111.4221}{{\tt
  1111.4221}}].

\bibitem{Alday:2009aq}
L.~F. Alday, D.~Gaiotto and Y.~Tachikawa, \emph{Liouville correlation functions
  from four-dimensional gauge theories},
  \href{http://dx.doi.org/10.1007/s11005-010-0369-5}{\emph{Lett. Math. Phys.}
  {\bf 91} (2010) 167--197} [\href{http://arxiv.org/abs/0906.3219}{{\tt
  0906.3219}}].

\bibitem{Wyllard:2009hg}
N.~Wyllard, \emph{{$A_{N-1}$ conformal Toda field theory correlation functions
  from conformal $\mathcal{N} = 2$ $\mathrm{SU}(N)$ quiver gauge theories}},
  \href{http://dx.doi.org/10.1088/1126-6708/2009/11/002}{\emph{JHEP} {\bf 11}
  (2009) 002} [\href{http://arxiv.org/abs/0907.2189}{{\tt 0907.2189}}].

\bibitem{Alday:2009fs}
L.~F. Alday, D.~Gaiotto, S.~Gukov, Y.~Tachikawa and H.~Verlinde, \emph{Loop and
  surface operators in {$\mathcal{N}=2$} gauge theory and {L}iouville modular
  geometry}, \href{http://dx.doi.org/10.1007/JHEP01(2010)113}{\emph{JHEP} {\bf
  01} (2010) 113} [\href{http://arxiv.org/abs/0909.0945}{{\tt 0909.0945}}].

\bibitem{Drukker:2009id}
N.~Drukker, J.~Gomis, T.~Okuda and J.~Teschner, \emph{Gauge theory loop
  operators and {L}iouville theory},
  \href{http://dx.doi.org/10.1007/JHEP02(2010)057}{\emph{JHEP} {\bf 02}
  (2010) 057} [\href{http://arxiv.org/abs/0909.1105}{{\tt 0909.1105}}].

\bibitem{Gomis:2010kv}
J.~Gomis and B.~Le~Floch, \emph{'t {H}ooft operators in gauge theory from
  {T}oda {CFT}}, \href{http://dx.doi.org/10.1007/JHEP11(2011)114}{\emph{JHEP} {\bf 11} (2011) 114} [\href{http://arxiv.org/abs/1008.4139}{{\tt
  1008.4139}}].

\bibitem{Verlinde:1988sn}
E.~Verlinde, \emph{Fusion rules and modular transformations in {$2$}{D}
  conformal field theory},
  \href{http://dx.doi.org/10.1016/0550-3213(88)90603-7}{\emph{Nucl. Phys. B}
  {\bf 300} (1988) 360--376}.

\bibitem{Costello:2013zra}
K.~Costello, \emph{Supersymmetric gauge theory and the {Y}angian},
  \href{http://arxiv.org/abs/1303.2632}{{\tt 1303.2632}}.

\bibitem{Costello:2017dso}
K.~Costello, E.~Witten and M.~Yamazaki, \emph{Gauge theory and integrability,
  {I}}, \href{http://dx.doi.org/10.4310/ICCM.2018.v6.n1.a6}{\emph{ICCM Not.}
  {\bf 6} (2018) 46--119} [\href{http://arxiv.org/abs/1709.09993}{{\tt
  1709.09993}}].

\bibitem{Costello:2018txb}
K.~Costello and J.~Yagi, \emph{Unification of integrability in supersymmetric
  gauge theories},  \href{http://arxiv.org/abs/1810.01970}{{\tt 1810.01970}}.

\bibitem{Felder:1994be}
G.~Felder, \emph{Elliptic quantum groups},  in \emph{X{I}th {I}nternational
  {C}ongress of {M}athematical {P}hysics ({P}aris, 1994)}, pp.~211--218, Int.
  Press, Cambridge, MA, 1995.
\newblock \href{http://arxiv.org/abs/hep-th/9412207}{{\tt hep-th/9412207}}.

\bibitem{Felder:1994pb}
G.~Felder, \emph{Conformal field theory and integrable systems associated to
  elliptic curves},  in \emph{Proceedings of the {I}nternational {C}ongress of
  {M}athematicians, {V}ol.\ 1, 2 ({Z}\"urich, 1994)}, pp.~1247--1255,
  Birkh\"auser, Basel, 1995.
\newblock \href{http://arxiv.org/abs/hep-th/9407154}{{\tt hep-th/9407154}}.

\bibitem{MR1645196}
P.~Etingof and A.~Varchenko, \emph{Solutions of the quantum dynamical
  {Y}ang-{B}axter equation and dynamical quantum groups},
  \href{http://dx.doi.org/10.1007/s002200050437}{\emph{Comm. Math. Phys.} {\bf
  196} (1998) 591--640} [\href{http://arxiv.org/abs/q-alg/9708015}{{\tt
  q-alg/9708015}}].

\bibitem{Gervais:1983ry}
J.-L. Gervais and A. Neveu, \emph{Novel triangle relation and absence of tachyons in Liouville string field theory},
  \href{http://dx.doi.org/10.1016/0550-3213(84)90469-3}{\emph{Nucl. Phys. B}
  {\bf 238} (1984) 125--141}.

\bibitem{Baxter:1971cr}
R.~J. Baxter, \emph{Eight-vertex model in lattice statistics},
  \href{http://dx.doi.org/10.1103/PhysRevLett.26.832}{\emph{Phys. Rev. Lett.}
  {\bf 26} (1971) 832--833}.

\bibitem{Baxter:1972hz}
R.~J. Baxter, \emph{Partition function of the eight-vertex lattice model},
  \href{http://dx.doi.org/10.1016/0003-4916(72)90335-1}{\emph{Ann. Phys.} {\bf
  70} (1972) 193--228}.

\bibitem{Belavin:1981ix}
A.~A. Belavin, \emph{Dynamical symmetry of integrable quantum systems},
  \href{http://dx.doi.org/10.1016/0550-3213(81)90414-4}{\emph{Nucl. Phys. B}
  {\bf 180} (1981) 189--200}.

\bibitem{Bazhanov:2010kz}
V.~V. Bazhanov and S.~M. Sergeev, \emph{A master solution of the quantum
  {Y}ang-{B}axter equation and classical discrete integrable equations},
  \href{http://dx.doi.org/10.4310/ATMP.2012.v16.n1.a3}{\emph{Adv. Theor. Math.
  Phys.} {\bf 16} (2012) 65--95} [\href{http://arxiv.org/abs/1006.0651}{{\tt
  1006.0651}}].

\bibitem{Bazhanov:2011mz}
V.~V. Bazhanov and S.~M. Sergeev, \emph{Elliptic gamma-function and multi-spin
  solutions of the {Y}ang-{B}axter equation},
  \href{http://dx.doi.org/10.1016/j.nuclphysb.2011.10.032}{\emph{Nucl. Phys. B}
  {\bf 856} (2012) 475--496} [\href{http://arxiv.org/abs/1106.5874}{{\tt
  1106.5874}}].

\bibitem{Gaiotto:2008cd}
D. Gaiotto, A. Neitzke and G.~W. Moore, \emph{Four-dimensional wall-crossing via three-dimensional field theory},
\href{http://dx.doi.org/10.1007/s00220-010-1071-2}{\emph{Comm. Math. Phys.} {\bf 299} (2010)
  163--224} [\href{http://arxiv.org/abs/0807.4723}{{\tt 0807.4723}}].

\bibitem{Brennan:2019hzm}
T.~D. Brennan and G.~W. Moore, \emph{Index-like theorems from line defect vevs},
\href{http://dx.doi.org/10.1007/jhep09(2019)073}{\emph{JHEP} {\bf 09} (2019)
  073} [\href{http://arxiv.org/abs/1903.08172}{{\tt 1903.08172}}].

\bibitem{Brennan:2018yuj}
T.~D. Brennan, A.~Dey and G.~W. Moore, \emph{On 't {H}ooft defects, monopole
  bubbling and supersymmetric quantum mechanics},
\href{http://dx.doi.org/10.1007/jhep09(2018)014}{\emph{JHEP} {\bf 09} (2018)
  014} [\href{http://arxiv.org/abs/1801.01986}{{\tt 1801.01986}}].

\bibitem{Brennan:2018moe}
T.~D. Brennan, \emph{Monopole bubbling via string theory},
  \href{http://dx.doi.org/10.1007/jhep11(2018)126}{\emph{JHEP} {\bf 11} (2018)
  126} [\href{http://arxiv.org/abs/1806.00024}{{\tt 1806.00024}}].

\bibitem{Brennan:2018rcn}
T.~D. Brennan, \emph{’t {H}ooft defects and wall crossing in {SQM}},
  \href{http://dx.doi.org/10.1007/jhep11(2019)173}{\emph{JHEP} {\bf 10} (2019)
  173} [\href{http://arxiv.org/abs/1810.07191}{{\tt 1810.07191}}].

\bibitem{Assel:2019iae}
B.~Assel and A.~Sciarappa, \emph{On monopole bubbling contributions to 't
  {H}ooft loops}, \href{http://dx.doi.org/10.1007/jhep05(2019)180}{\emph{JHEP}
  {\bf 05} (2019) 180} [\href{http://arxiv.org/abs/1903.00376}{{\tt
  1903.00376}}].

\bibitem{Ruijsenaars:1986pp}
S.~N.~M. Ruijsenaars, \emph{Complete integrability of relativistic
  {C}alogero--{M}oser systems and elliptic function identities},
\href{https://doi.org/10.1007/BF01207363}{\emph{Comm.
  Math. Phys.} {\bf 110} (1987) 191--213}.

\bibitem{Gaiotto:2009we}
D.~Gaiotto, \emph{{$N=2$} dualities},
  \href{http://dx.doi.org/10.1007/JHEP08(2012)034}{\emph{JHEP} {\bf 08} (2012)
  034} [\href{http://arxiv.org/abs/0904.2715}{{\tt 0904.2715}}].

\bibitem{Gaiotto:2009hg}
D.~Gaiotto, G.~W. Moore and A.~Neitzke, \emph{Wall-crossing, {H}itchin systems,
  and the {WKB} approximation},
  \href{http://dx.doi.org/10.1016/j.aim.2012.09.027}{\emph{Adv. Math.} {\bf
  234} (2013) 239--403} [\href{http://arxiv.org/abs/0907.3987}{{\tt
  0907.3987}}].

\bibitem{Witten:1997sc}
E.~Witten, \emph{Solutions of four-dimensional field theories via
  {$M$}-theory},
  \href{http://dx.doi.org/10.1016/S0550-3213(97)00416-1}{\emph{Nucl. Phys. B}
  {\bf 500} (1997) 3--42} [\href{http://arxiv.org/abs/hep-th/9703166}{{\tt
  hep-th/9703166}}].

\bibitem{Yagi:2014toa}
J.~Yagi, \emph{{$\Omega$}-deformation and quantization},
  \href{http://dx.doi.org/10.1007/JHEP08(2014)112}{\emph{JHEP} {\bf 08} (2014)
  112} [\href{http://arxiv.org/abs/1405.6714}{{\tt 1405.6714}}].

\bibitem{Hellerman:2011mv}
S.~Hellerman, D.~Orlando and S.~Reffert, \emph{String theory of the Omega deformation},
  \href{http://dx.doi.org/10.1007/JHEP01(2012)148}{\emph{JHEP} {\bf 01} (2012)
  148} [\href{http://arxiv.org/abs/1106.0279}{{\tt 1106.0279}}].

\bibitem{Nekrasov:2002qd}
  N.~Nekrasov, \emph{Seiberg--Witten prepotential from instanton counting},
  \href{http://dx.doi.org/10.4310/ATMP.2003.v7.n5.a4}{\emph{Adv. Theor. Math. Phys.} {\bf 7} (2003) 831--864} [\href{http://arxiv.org/abs/hep-th/0206161}{{\tt hep-th/0206161}}].


\end{thebibliography}
\end{document}